\documentclass[10pt,dvips,a4paper]{article}
\usepackage[pdftex]{graphicx}
\usepackage{amssymb}
\usepackage{amsmath}
\usepackage{setspace}
\usepackage[round]{natbib}
\usepackage[pdftex]{graphicx,xcolor}

\newcommand{\ignore}[1]{}
\newcommand{\optional}[1]{}
\newcommand{\dro}{\bar{r}_{\rm o}}
\newcommand{\dr}{\bar{r}}
\newcommand{\du}{\bar{u}}
\newcommand{\dPfi}{\bar{P}_{\rm fi}}
\newcommand{\dPf}{\bar{P}_{\rm f}}

\renewcommand{\title}{On a 2D hydro-mechanical lattice approach for modelling hydraulic fracture}

\begin{document}

\begin{center} \begin{LARGE} \textbf{\title} \end{LARGE} \end{center}

\begin{center}  P. Grassl$^{\mbox{1}*}$, C. Fahy$^{\mbox{1}}$, D. Gallipoli$^{\mbox{2}}$, S. J. Wheeler$^{\mbox{1}}$  \end{center}

$^{1}$School of Engineering, University of Glasgow, Glasgow, UK\\
$^{2}$Universit\'{e} de Pau et des Pays de l'Adour, Laboratoire SIAME, France \\

$^{*}$ corresponding author: Email: peter.grassl@glasgow.ac.uk

 
\section*{Abstract}
A 2D lattice approach to describe hydraulic fracturing is presented. 
The interaction of fluid pressure and mechanical response is described by Biot's theory.
The lattice model is applied to the analysis of a thick-walled cylinder, for which an analytical solution for the elastic response is derived. The numerical results obtained with the lattice model agree well with the analytical solution.
Furthermore, the coupled lattice approach is applied to the fracture analysis of the thick-walled cylinder.
It is shown that the proposed lattice approach provides results that are independent of the mesh size. Moreover, a strong geometrical size effect on nominal strength is observed which lies between analytically derived lower and upper bounds. This size effect decreases with increasing Biot's coefficient.

\section{Introduction}

The modelling of the coupling of water pressure and fracture is important for, e.g., hydraulic fracturing for oil and gas extraction \citep{BazSalChau14}, failure of flood defense embankments, and earth and concrete dams \citep{SloSao00}. Furthermore, many naturally occuring phenomena in tectonophysics can be explained by fracturing induced by fluid pressure: injection of sills \citep{Gou05} and clastic dykes \citep{Mee09}.

The aim of the present work is to propose a coupled hydro-mechanical lattice approach for modelling fracture in saturated porous materials. The model is based on a combination of previously developed lattice techniques for the mechanical response \citep{GraJir10} and mass transport \citep{Gra09}.  
The theories of \citet{Bio41} and \citet{Ter25} are used to couple the transport and mechanical lattice model.
The lattice models used belong to the group of discrete approaches. Another group of discrete approaches are particle models.  
In particle models, the arrangement of particles evolve such that neighbours of particles change during the analysis.
On the other hand, for lattice models the connectivity of elements does not change during the analysis.
These lattice models are very suitable for describing fracture initiation and propagation in quasi-brittle materials such as concrete and rocks \citep{ZubBaz87,BazTabKazPij90,JirBaz95,DelPijRou96,BolSai98,YipMohBol05}.
Lattice models have been used to describe the interaction of the mechanical response with other physical processes, such as the modelling of the influence of moisture transport on drying shrinkage \citep{BolBer04,AsaHouBir14} or cracking on mass transport \citep{Gra09,WanUed11a,SavPacSch13}.

The model performance is assessed by two studies.
In the first study, the model is applied to the analysis of an elastic uncracked thick-walled cylinder subjected to an inner fluid pressure leading to steady state flow for varying Biot's coefficient.
The model results are compared to the corresponding analytical solution derived in the present study.
The analytical derivations are related to the work presented in \citet{ShaElw70}, which considered the response of the thick-walled cylinder for the case of an inner fluid pressure and Terzaghi's effective stress, which is one of the limiting cases of the present study for varying Biot's coefficients.
Related work on fracture initiation of saturated thick-walled cylinders was also presented in \citet{RicCle76}.
The analytical derivations have also some similarities to the work on cavity expansion proposed originally in \citet{BisHilMot45} and later for geotechnical applications in \citet{YuHou91} and \citet{Yu00}. The elastic solution for this type of method has also been presented in textbooks such as \citet{TimGoo87}.
However, in cavity expansion approaches the load in the inside of the thick-walled cylinder (cylindrical cavity) is applied in the form of a mechanical and not a fluid pressure, as it is done here.

In the second study, the analysis of the thick-walled cylinder is extended to fluid pressure driven fracture. 
The influence of the Biot's coefficient and the size of the lattice elements on the load capacity and fracture patterns are investigated. 
Furthermore, the effect of size of the thick-walled cylinder on the normalised load capacity is studied \citep{Baz01}.
The numerical results are compared to upper and lower bounds obtained from the analytical solution.

\section{Lattice Model}\label{sec:LatticeModel}
\begin{figure}
\begin{tabular}{cc}
\includegraphics[height=5.5cm]{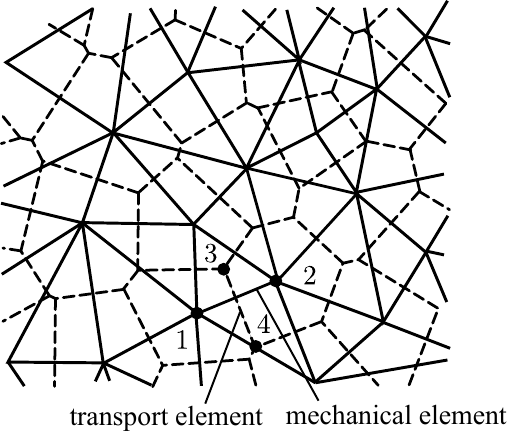} & \includegraphics[height=5.5cm]{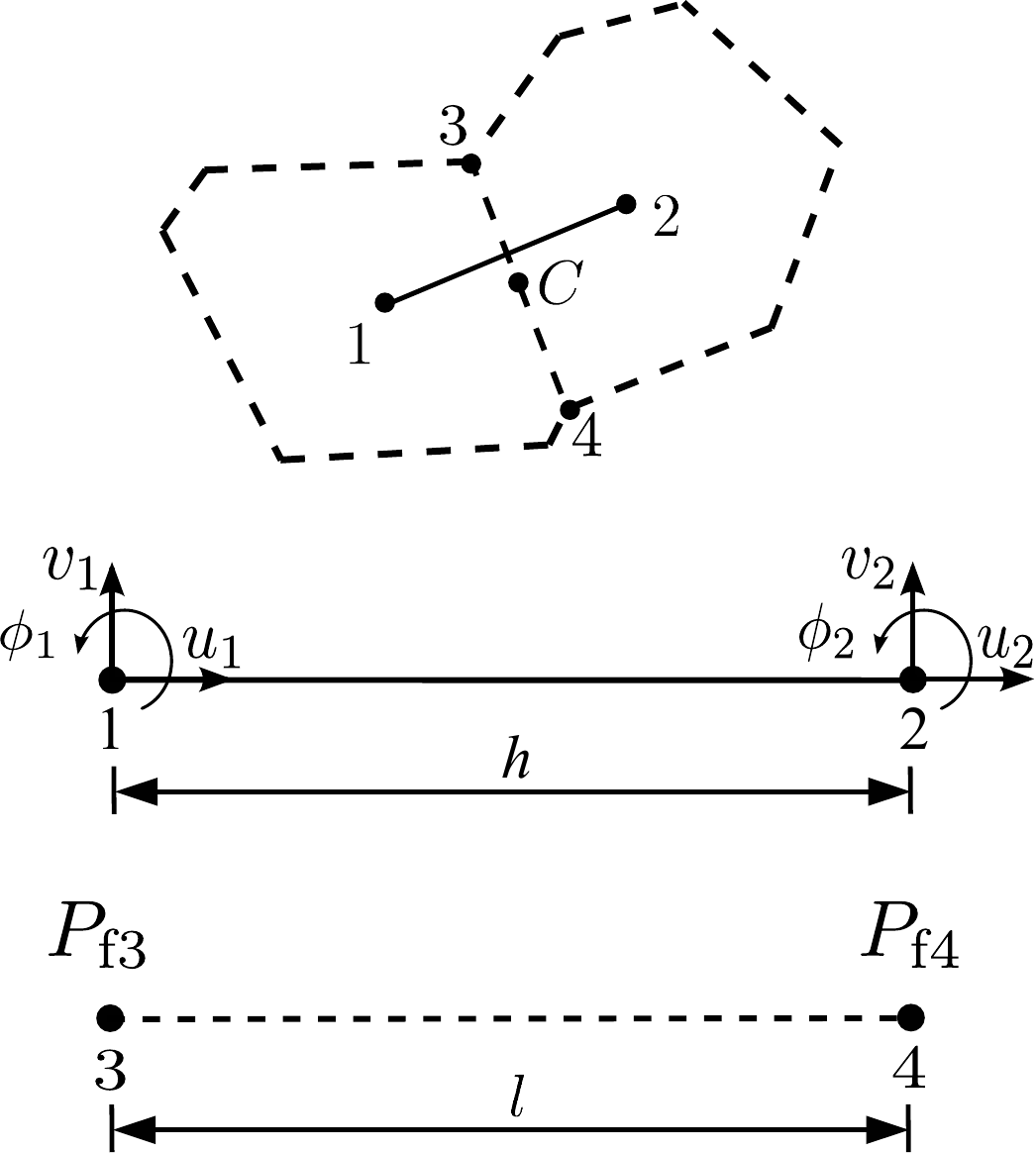}
\end{tabular}
\caption{Structure of the lattices: (a) Dual mechanical and transport lattices, (b) single mechanical and transport elements.}
\label{fig:dualMesh}
\end{figure}

The fluid transport and the mechanical response within a 2D domain are modelled by lattices based on one-dimensional transport and mechanical elements. 
A dual Delaunay and Voronoi tessellation \citep{Aur91}, based on a set of randomly located vertices, is used to discretise the domain (Figure \ref{fig:dualMesh}(a)). 
The Delaunay tessellation results in triangles and the Voronoi tessellation in polygons.
The random placement of vertices is performed by a trial and error approach enforcing a minimum distance $d_{\rm min}$ between the vertices.
Random points are placed until a maximum number of iterations is required to place an additional point. 
At this stage, the specimen is considered to be fully saturated by random points.
Whereas the minimum distance between nodes is enforced, the average and maximum length of the Delaunay edges is only influenced by $d_{\rm min}$ but not directly prescribed.
The smaller $d_{\rm min}$ is, the smaller is also the mean and maximum length of Delaunay edges.
The minimum length of Voronoi edges is not enforced.

The mechanical elements are placed along the edges of the Delaunay triangles and their mid-cross section geometry is determined by the lengths of the corresponding Voronoi edges. 
The one-dimensional transport elements are placed along the edges of the Voronoi polygons and their cross-sectional properties are determined from the corresponding lengths of Delaunay triangle edges. 
The coupling of both lattices is enforced at the point $C$, located midway along the Voronoi edge, in Figure \ref{fig:dualMesh}(b) where the constitutive response for both the mechanical and transport approach is available. 
In Figure \ref{fig:dualMesh}(b), $u$, $v$ and $\phi$ are nodal degrees of freedom of the mechanical model and $P_{\rm f}$ is the nodal unknown (fluid pressure) of the transport model. 

\subsection{Mechanical Model}\label{subsec:MechLaw}

The equations for the mechanical lattice element have been discussed in detail in \citet{BolSai98} and \citet{GraGreSol12}.
Here, only the constitutive model of the mechanical part is discussed. 
The stress of the mechanical model, which enters the equilibrium condition, is split into an effective stress carried by the solid and a portion of fluid pressure in the pores by applying Biot's theory of porous solids \citep{Bio41,Cou11} as
\begin{equation}\label{eq:StressStrain}
\boldsymbol{\sigma}= \boldsymbol{\sigma}^{\rm m} + b \boldsymbol{\sigma}^{\rm f}
\end{equation}
where $\boldsymbol{\sigma}=\left\{\sigma_{\rm{n}},\sigma_{\rm{s}},\sigma_{\rm{\phi}}\right\}^{T}$ is the stress vector (comprising of normal, shear and rotational stresses), $\boldsymbol{\sigma}^{\rm m}=\left\{\sigma^{\rm m}_{\rm{n}},\sigma^{\rm m}_{\rm{s}},\sigma^{\rm m}_{\rm{\phi}}\right\}^{T}$ is the effective stress vector, $\boldsymbol{\sigma}^{\rm f}=\left\{P_{\rm f},0,0\right\}^{T}$ is the fluid pressure vector (see also section~\ref{subsec:TransLaw}) and $b$ is Biot's coefficient. For the  pore fluid pressure $P_{\rm f}$, the sign convention is tension positive.
For $b=0$, the total stress is equal to the effective stress, so that the fluid pressure has no influence on the mechanical response.
For $b = 1$, the full amount of fluid pressure is added to the effective stress, which corresponds to Terzaghi's principle of effective stress \citep{Ter25}.
The choice of the value of $b$ depends on the type of material to be modelled \citep{Cou11}.

For the effective stress, an elastic-damage model is used to describe the response of the lattice elements. The stress-strain law for this part is
\begin{equation}\label{eq:stressStrainMech}
\boldsymbol{\sigma}^{\rm m}=(1-\omega) \mathbf{D} \boldsymbol{\varepsilon}
\end{equation}
where $\mathbf{D}$ is the elastic stiffness, $\boldsymbol{\varepsilon}=\left\{\varepsilon_{\rm{n}},\varepsilon_{\rm{s}},\varepsilon_{\rm{\phi}}\right\}^{T}$ is the strain vector (comprising of normal, shear and rotational strains) and $\omega$ is the damage parameter.  
The elastic stiffness is defined as
\begin{equation}
\boldsymbol{\rm D} = \left[ \begin{array}{ccc}
E & 0 & 0 \\
0 & \gamma E & 0 \\
0 & 0 & E \end{array} \right]
\end{equation}
and depends on the model parameters $E$ and $\gamma$. 
For plane stress analysis and a regular lattice of equilateral triangles, these model parameters are related to the continuum Young's modulus $E_{\rm c}$ and Poisson's ratio $\nu$ as
\begin{equation}
\gamma=\dfrac{1-3\nu}{\nu+1}
\label{eqn:gamma}
\end{equation}
and
\begin{equation}
E=\dfrac{E_{\mathrm{c}}}{1-\nu}
\label{eqn:E}
\end{equation}
as derived in \citet{GriMus01}.
For the irregular lattice used in this study, the expressions in \eqref{eqn:gamma} and \eqref{eqn:E} are used as approximations.
For $\nu > 1/3$, $\gamma <0$, which corresponds to a negative shear stiffness, which is considered not to be physical. Therefore, for the present lattice model the maximum value of the Poisson's ratio is theoretically limited to $1/3$. Pratically, the present random discrete model provides accurate results only for $\nu \leq 0.2$. Therefore, $\nu = 0.2$ is the upper limit used for the numerical study in Section~\ref{sec:analysis}.

The damage parameter $\omega$ is a function of a history variable $\kappa$, which is determined by the loading function
\begin{equation} \label{eq:loadingFunc}
f(\boldsymbol{\varepsilon},\kappa) = \varepsilon_{\rm eq} \left( \boldsymbol{\varepsilon} \right) - \kappa 
\end{equation}
and the loading-unloading conditions 
\begin{equation}\label{eq:loadunload}
f \leq 0 \mbox{,} \hspace{0.5cm} \dot{\kappa} \geq 0 \mbox{,} \hspace{0.5cm} \dot{\kappa} f = 0
\end{equation}

The equivalent strain $\varepsilon_{\rm eq}$ in (\ref{eq:loadingFunc}) is defined as
\begin{equation} \label{eq:equiv}
\varepsilon_{\rm eq}(\varepsilon_{\rm n},\varepsilon_{\rm s}) = \dfrac{1}{2} \varepsilon_0 \left( 1-c \right) + \sqrt{\left( \dfrac{1}{2} \varepsilon_0 (c-1) + \varepsilon_{\rm n}\right)^2 + \dfrac{ c \gamma^2 \varepsilon_{\rm s}^2}{q^2}}
\end{equation} 
where $\varepsilon_0$, $c$ and $q$ are model parameters, which are directly related to the strength and stiffness of the equivalent continuum of the lattice elements.
The present equivalent strain definition depends only on the first two strain components, namely $\varepsilon_{\rm n}$ and $\varepsilon_{\rm s}$ \citep{GraJir10}. 
However, all three elastic stress components in (\ref{eq:StressStrain}) are reduced by the damage parameter $\omega$.
\begin{figure}
\begin{center}
\begin{tabular}{cc}
\includegraphics[width=7cm]{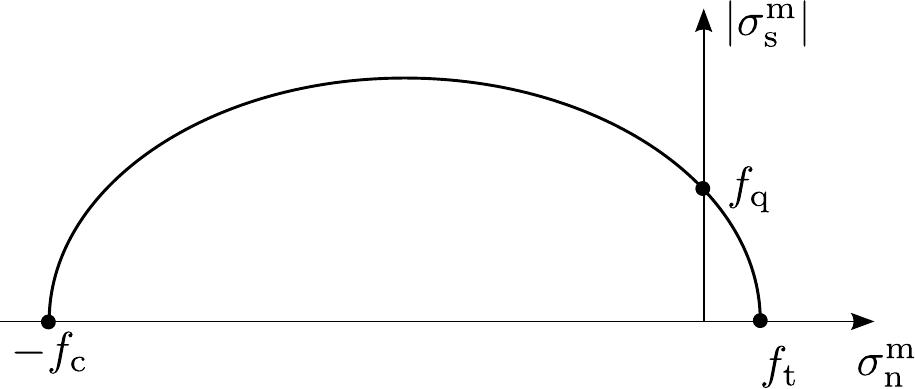} & \includegraphics[width=5cm]{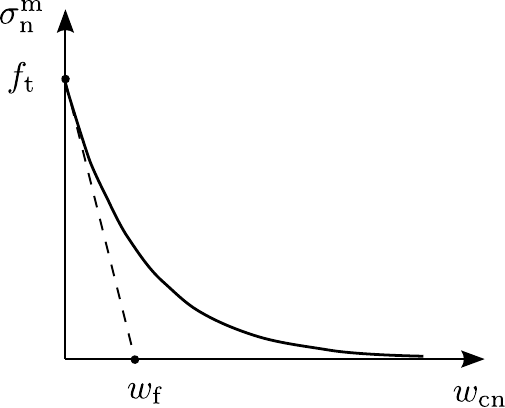}\\
(a) & (b)
\end{tabular}
\caption{(a) Elliptic strength envelope in mechanical stress space and (b) exponential stress crack opening curve.}
\label{fig:envelope}
\end{center}
\end{figure}
This equivalent strain definition results in an elliptic strength envelope shown in Fig.~\ref{fig:envelope}.
For pure tension, the stress is limited by the tensile strength $f_{\rm t} = E \varepsilon_{0}$.
For pure shear and pure compression, the stress is limited by the shear strength $f_{\rm q} = q f_{\rm t}$ and the compressive strength $f_{\rm c} = c f_{\rm t}$, respectively.

The expression for the damage parameter $\omega$ is derived by considering the case of pure tension. 
The softening curve of the stress-strain response in pure tension is chosen as
\begin{equation} \label{eq:exp}
\sigma^{\rm m}_{\rm n} = f_{\rm t} \exp \left(-\dfrac{w_{\rm cn}}{w_{\rm f}}\right)
\end{equation}
where $w_{\rm cn} = \omega h \varepsilon_{\rm n}$ is considered as a crack opening in monotonic pure tension (Fig.~\ref{fig:envelope}(b)).
Here, $h$ is the length of the mechanical element shown in Fig.~\ref{fig:dualMesh}(b).
The stress-strain relation in pure tension can also be expressed by (\ref{eq:stressStrainMech}) in terms of the damage variable as
\begin{equation}\label{eq:uni}
\sigma^{\rm m}_{\rm n} =  \left(1-\omega \right) E \varepsilon_{\rm n}
\end{equation}
Comparing the right-hand sides of (\ref{eq:exp}) and (\ref{eq:uni}), 
and replacing $\varepsilon_{\rm n}$ by $\kappa$, we obtain the equation
\begin{equation}\label{eq:omega}
\left(1-\omega \right) \kappa = \varepsilon_0\exp \left(-\dfrac{\omega h \kappa}{w_{\rm f}}\right)
\end{equation}
from which the damage parameter $\omega$ can be determined iteratively using, for instance, a Newton-Raphson method.
In (\ref{eq:omega}), the history variable $\kappa$ is the maximum equivalent strain reached in the history of the material point, which in turn is, in general, a function of $\varepsilon_{\rm n}$ and $\varepsilon_{\rm s}$.
Thus, the damage parameter determined from (\ref{eq:omega}) will only result in the exponential stress crack opening curve in (\ref{eq:exp}), if the element is subjected to pure tension.
Parameter $w_{\rm f}$ determines the initial slope of the softening curve and is related to the meso-level fracture energy $G_{\rm F} = f_{\rm t} w_{\rm f}$ for pure tensile loading, which corresponds to the total area under the stress crack opening curve in Fig.~\ref{fig:envelope}(b).
For combinations of $\varepsilon_{\rm n}$ and $\varepsilon_{\rm s}$, the energy disspated by an element will differ from $G_{\rm F}$.
The model parameters for the mechanical part are $E$ and $\gamma$ for the elastic response, and $\varepsilon_0$, $q$, $c$ and $w_{\rm f}$ for the inelastic response.

\subsection{Transport Model}\label{subsec:TransLaw}

The lattice model for the fluid transport is based on several idealisations.
Fluid pressures applied at the boundary of the model are assumed to result in steady-state conditions. Thus, any transient effects due to changes of the boundary pressure and changes of the transport properties are disregarded. Instead, the final state, for which equilibrium with the hydraulic boundary conditions hold, is considered. This idealised approach is well suited for applications in which the change of boundary pressure and crack propagation occur slowly.  
Furthermore, it is assumed that the transport properties are independent of the mechanical response and the fluid is incompressible. This is an idealisation of many applications in which, for instance, mechanical loading might introduce changes in hydraulic conductivity and fluid compressibility could be significant. Our motivation for this idealisation was to be able to create a modelling framework with a small number of input parameters, for which we can identify dominant trends of the effect of fluid pressure on the mechanical response by parametric study, and compare the numerical results to analytical solutions.

The transport of an incompressible fluid through a saturated porous medium in steady-state conditions is governed by the conservation of fluid mass in the form
\begin{equation} \label{eq:conservation}
\rho \mbox{div}(\mathbf{q}) = 0
\end{equation}
where $\rho$ is the constant fluid density and $\mathbf{q}$ is the flux vector (i.e. the vector of flow rate per unit area of porous medium).
Under laminar flow conditions, Darcy's law relates the flux vector to the gradient of the pore fluid pressure as
\begin{equation} \label{eq:flux}
\mathbf{q} = k \mbox{grad} P_{\rm f}
\end{equation}
where $k$ is the hydraulic conductivity, $P_{\rm f}$ is the pore fluid pressure (tension positive).
In (\ref{eq:flux}), it is assumed that the effect of gravity is negligible. This is acceptable, if the magnitude of the gradient of the pore fluid pressure is significantly larger than the gradient of the gravitational potential energy, or if fluid flow takes place in a horizontal plane at a given elevation.
By substituting (\ref{eq:flux}) into (\ref{eq:conservation}), and assuming constant conductivity $k$, the differential equation for the transport model in the form
\begin{equation} \label{eq:fullCons}
 \rho k \mbox{div}\left(\mbox{grad} P_{\rm f}\right) = 0
\end{equation}
is obtained.

Conditions are imposed on the boundary $\Gamma$ of the material domain, either as prescribed values of fluid pressure (on sub-boundary $\Gamma_1$) or as prescribed values of the flux in the direction perpendicular to the boundary (on sub-boundary $\Gamma_2$). The latter condition can then be related to the gradient of the fluid pressure through Darcy's law. These two boundary conditions can be formalised as
\begin{equation}
P_{\rm f}=g(\mathbf{x})\;{\rm on}\;\Gamma_{\rm{1}}\quad \mbox{and}\quad \dfrac{\partial P_{\rm f}}{\partial\mathbf{n}}=f(\rm{\boldsymbol{\rm x}})\;\mbox{on}\;\Gamma_{2} \quad \mbox{with} \quad \Gamma = \Gamma_1 \cup \Gamma_2   
\end{equation}
where $\mathbf{n}$ denotes the direction normal to the boundary and $g(\mathbf{x})$ and $f(\mathbf{x})$ are scalar functions of the spatial coordinate vector $\mathbf{x}$.

The differential equation for mass transport in (\ref{eq:fullCons}) is modeled by a lattice of one-dimensional transport elements. The discrete form of one of these elements is 
\begin{equation}\label{eq:discretePDF}
\boldsymbol{\alpha}_{\rm e} \mathbf{P}_{\rm f} = \mathbf{f}_{\rm e}
\end{equation}
where $\mathbf{P}_{\rm f}$ is a vector containing the nodal values of the fluid pressure, $\boldsymbol{\alpha}_{\rm e}$ is the conductivity matrix and $\mathbf{f}_{\rm e}$ is the nodal flow rate vector.
The conductivity matrix is defined as 
\begin{equation}\label{eq:conMatrix}
\boldsymbol{\alpha}_{\rm e} = \rho \dfrac{A}{l} k \begin{pmatrix} 1 & -1\\ -1 & 1 \end{pmatrix}
\end{equation}
where $A = ht$ is the cross-sectional area, and $h$ and $l$ are defined in Figure~\ref{fig:dualMesh}(b).
Here, $t$ is the out-of-plane thickness of the element assuming a rectangular cross-section.
The model input parameters for the transport part are the density $\rho$ and the hydraulic conductivity $k$.

\subsection{Coupling of boundary conditions for the mechanical and transport problem} \label{sec:coupling}
The aim of this study is to model fracture caused by fluid pressure. For this type of loading, the total stress at the boundary is equal to the fluid pressure. Therefore, the mechanical (effective stress) at the boundary is zero. 
At the boundaries, two approaches were used to couple the boundary condition of the mechanical and transport model.

\begin{figure}
\begin{center}
\begin{tabular}{cc}
\includegraphics[height=4.5cm]{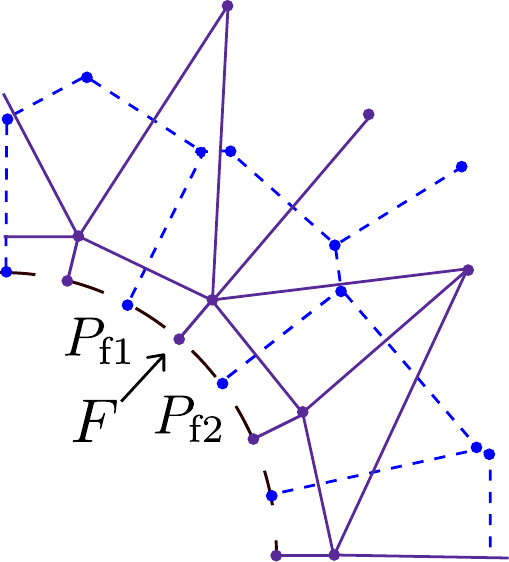} & \includegraphics[height=4.5cm]{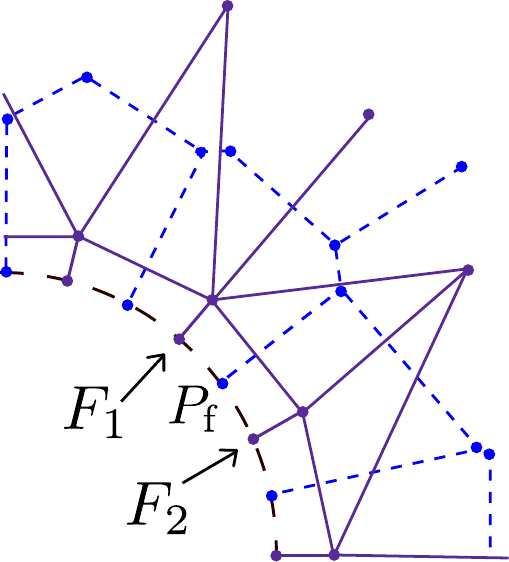}\\
(a) & (b)
\end{tabular}
\end{center}
\caption{Coupled boundary conditions: (a) Approach 1 - forces on structural nodes derived from applied pressures; and (b) Approach 2 - pressures at transport nodes derived from reactive forces associated with applied displacements.}
\label{fig:coupledBoundary}
\end{figure}

In Figure \ref{fig:coupledBoundary}, the two approaches are illustrated by showing the detail of the lattice at the boundary, which was generated by strategically placing nodes of the mechanical lattice along the boundary and manually moving nodes of the transport lattice, so that they too lie on the boundary, which is in this example circular.

In approach 1, the fluid pressure was prescribed at the nodes of the transport problem, representing the boundary. 
This fluid pressure was then used to calculate forces normal to the boundary of the mechanical problem by multiplying them with the cross-sectional area of the element. 
A schematic illustration of this transfer of fluid pressures $P_{\rm f1}$ and $P_{\rm f2}$  to radial boundary force $F$ is shown in Figure~\ref{fig:coupledBoundary}(a). 
This approach for the coupling of boundary conditions was used for the first part of the analysis in Section~\ref{sec:analysis}, in which the model was compared to the analytical solution for an elastic material response. 

For approach 2, a displacement normal to the boundary was prescribed in the mechanical model at the boundary of the specimen. 
The resulting reaction forces normal to the boundary were then converted into a pressure by dividing them by the cross-sectional area of the element. This pressure is applied to the boundary nodes of the transport model.
This boundary condition was used for the fracture analyses in Section~\ref{sec:analysis}, since it allowed an increase of the radial displacement to be obtained even for decreasing fluid pressure.
A schematic illustration of this type of coupled boundary condition, where reaction forces $F_1$ and $F_2$ are transferred to the fluid pressure $P_{\rm f}$, is shown in Figure~\ref{fig:coupledBoundary}(b).
For both types of boundary conditions, the integral of the fluid pressure at the transport boundary nodes along the boundary is equal to the sum of the forces at the mechanical nodes along the boundary.

\section{Lattice analyses of the elastic response of a thick-walled cylinder}\label{sec:analysis}
\begin{figure}
\begin{center}
\includegraphics[height=5.1cm]{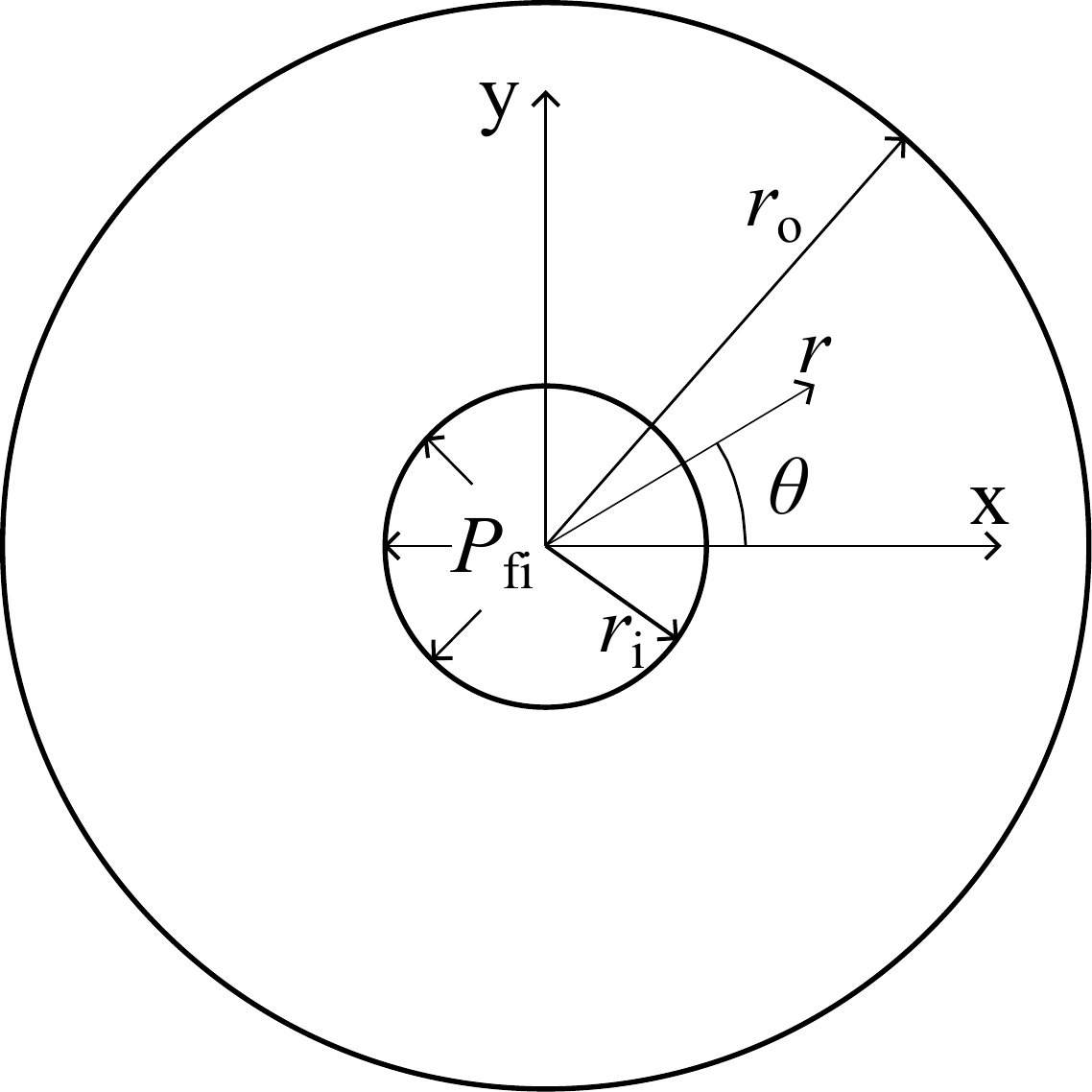}
\end{center}
\caption{Geometry of the thick-walled cylinder.}
 \label{fig:cylinder}
\end{figure}

\begin{figure}
\begin{center}
\begin{tabular}{cc}
\includegraphics[height=5.4cm]{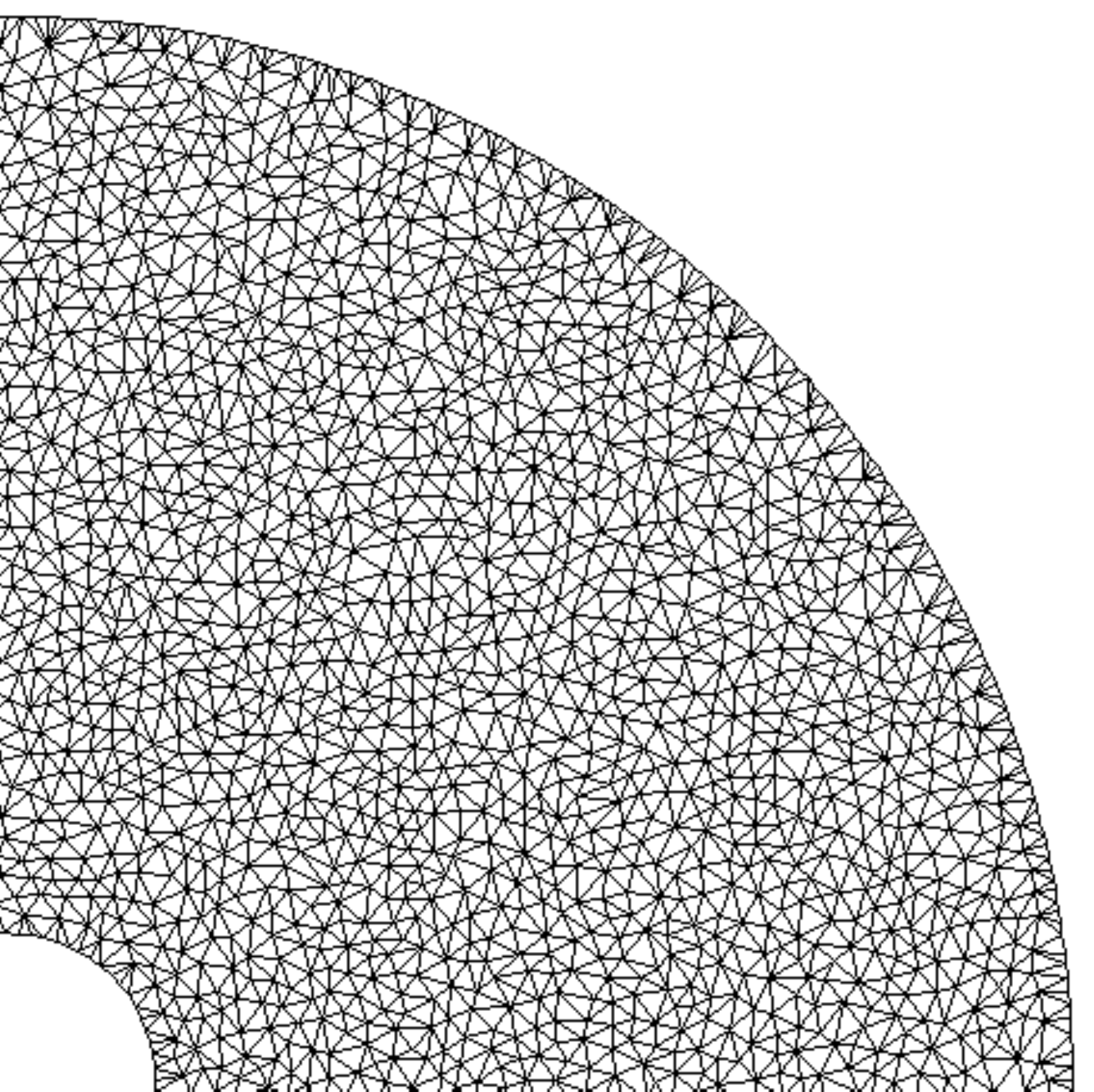} & \includegraphics[height=5.4cm]{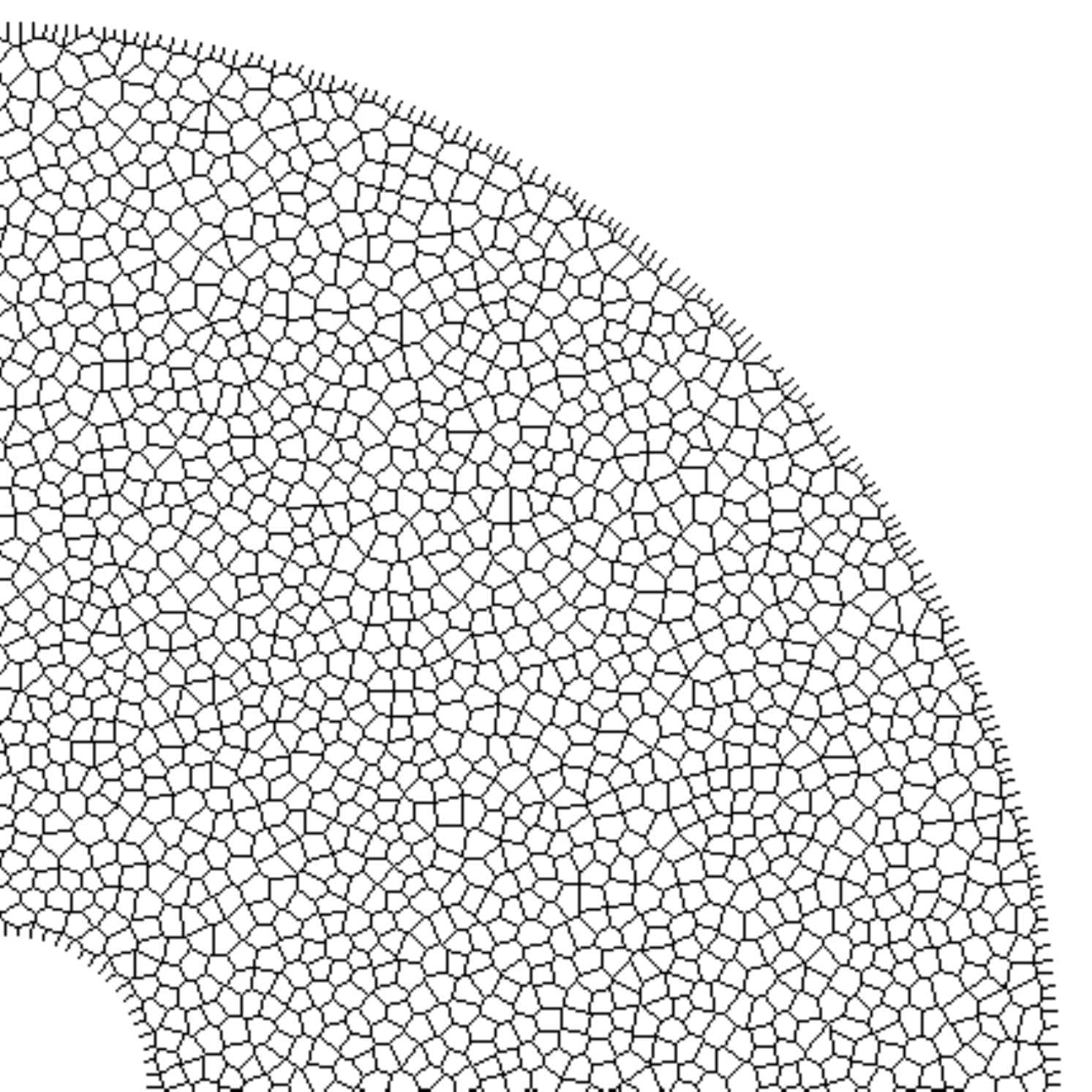}\\
(a) & (b)
\end{tabular}
\end{center}
\caption{A quarter of the lattice of (a) mechanical and (b) transport elements.}
 \label{fig:mesh}
\end{figure}

The capability of the coupled lattice approach to describe the interaction of flow and mechanical response is demonstrated by analysing the fluid pressure and elastic radial displacement distribution in a thick-walled cylinder in Figure~\ref{fig:cylinder} subjected to an inner fluid pressure, $P_{\rm fi}$.
A compressive internal fluid pressure ($P_{\rm fi}<0$) leads to expansion of the cylindrical cavity, whereas a tensile internal fluid pressure ($P_{\rm fi}>0$) leads to a contraction of the cavity.
For analysing the thick-walled cylinder, a cylindrical coordinate system is used with the radial, circumferential and out-of-plane coordinate axis denoted as $r$, $\theta$ and $z$, respectively. Plane stress conditions are assumed in the $z$ direction. The radial displacement is denoted as $u$. 
All variables of the problem are presented in dimensionless form. 
Variables of dimension length are normalised by the inner radius $r_{\rm i}$ (Figure~\ref{fig:cylinder}).
All variables of dimension pressure are normalised by the Young's modulus $E_{\rm c}$, which is through (\ref{eqn:E}) related to the model parameter $E$.
The geometry of the thick-walled cylinder is defined by the inner and outer radii $r_{\rm i}$ and $r_{\rm o}$, respectively.
Here, the ratio of the two radii was chosen as $\dro = \dfrac{r_{\rm o}}{r_{\rm i}} = 7.25$.
The discretisation approach presented in Section~\ref{sec:LatticeModel} is used to generate the lattice for the mechanical and transport model.
The minimum distance used to place random vertices for the Delaunay triangulation was chosen as $\bar{d}_{\rm min} = \dfrac{d_{\rm{min}}}{r_{\rm i}} = 0.123$.
A quarter of the mechanical and transport lattices are shown in Figure~\ref{fig:mesh}(a)~and~(b), respectively.
The numerical results are compared to an analytical solution presented in the Appendix.
The influence of Biot's coefficients of $b = 0$, $0.5$ and $1$ and Poisson's ratios of $\nu = 0$, $0.1$ and $0.2$ are investigated. 
For the elastic analyses, the transport and mechanical boundary conditions were coupled by Approach 1 described in Section~\ref{sec:coupling}. In this approach, the analysis is controlled by the fluid pressure applied at the inner boundary of the cylinder.
For comparing the lattice results with the analytical results, the vertices were divided into groups with respect to their radial coordinate. 
For each group of vertices the mean of the radial coordinate and unknown are presented. 
This technique allows for a clearer comparison of computational and analytical results.
However, it also potentially hides any local fluctuations, which do not influence the means. 

The elastic model parameters are varied to obtain different Poisson's ratios (see~(\ref{eqn:E})~and~(\ref{eqn:gamma})).
All the results are normalised so that they are independent of the value of $E$ used.
The results presented are independent of the hydraulic conductivity $k$ and fluid density $\rho$.

\begin{figure}
\begin{center}
  \includegraphics[width=12.cm]{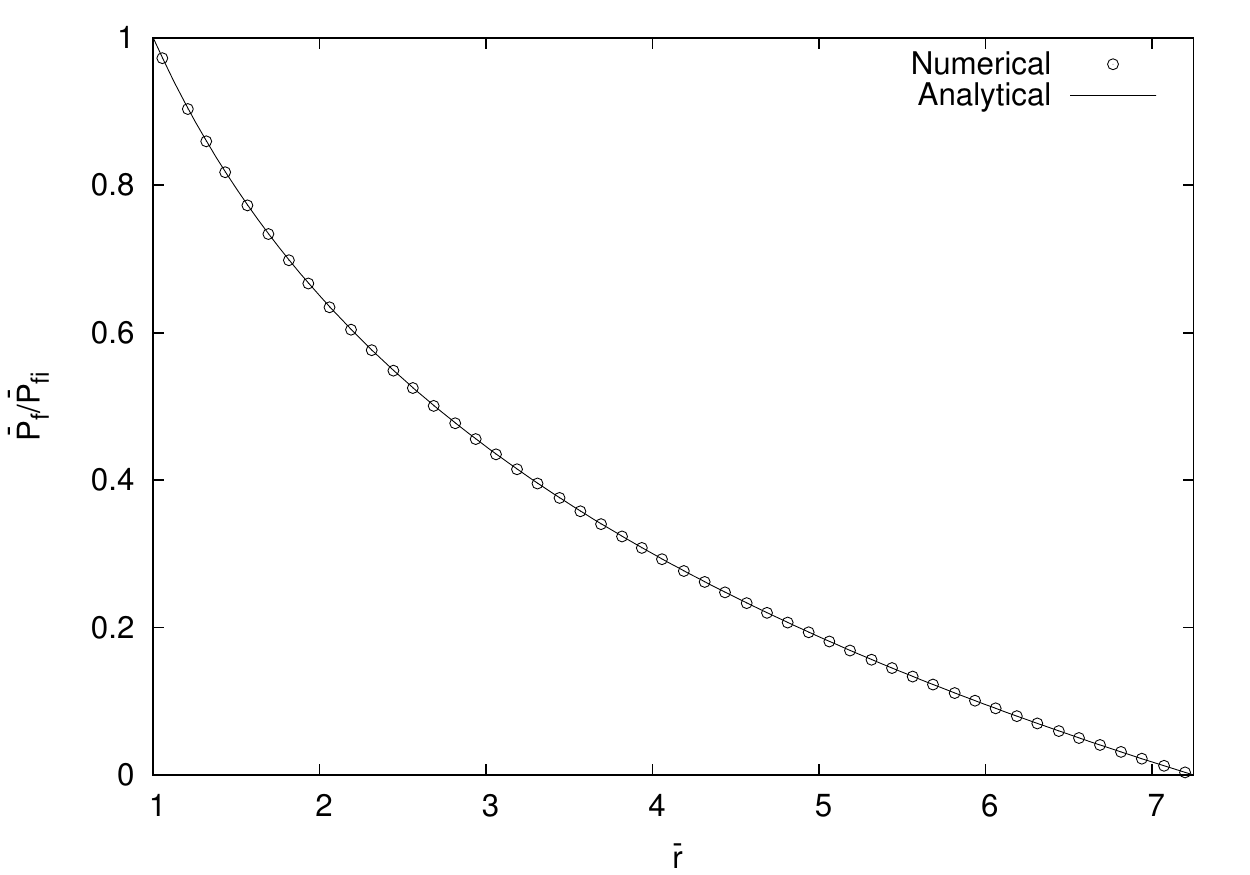}
\end{center}
 \caption{Ratio of normalised fluid pressure and normalised internal fluid pressure versus normalised radial coordinate for lattice model and analytical expression.}
\label{fig:pressure}
\end{figure}

In Figure~\ref{fig:pressure}, the fluid pressure distribution of the lattice model is compared to the analytical expression in (\ref{eq:dimFlow}). 
Here, the normalised variables are $\bar{r} = r/r_{\rm i}$, $\dPf = P_{\rm f}/E_{\rm c}$ and $\dPfi = P_{\rm fi}/E_{\rm c}$.
There is a very close agreement between lattice results and the analytical expression. This capability of the transport lattice model to reproduce analytical solutions has been previously reported for this type of tessellation for other problems in \citet{Gra09}.

\begin{figure}
\begin{center}
\begin{tabular}{cc}
(a) &  \includegraphics[width=10.cm]{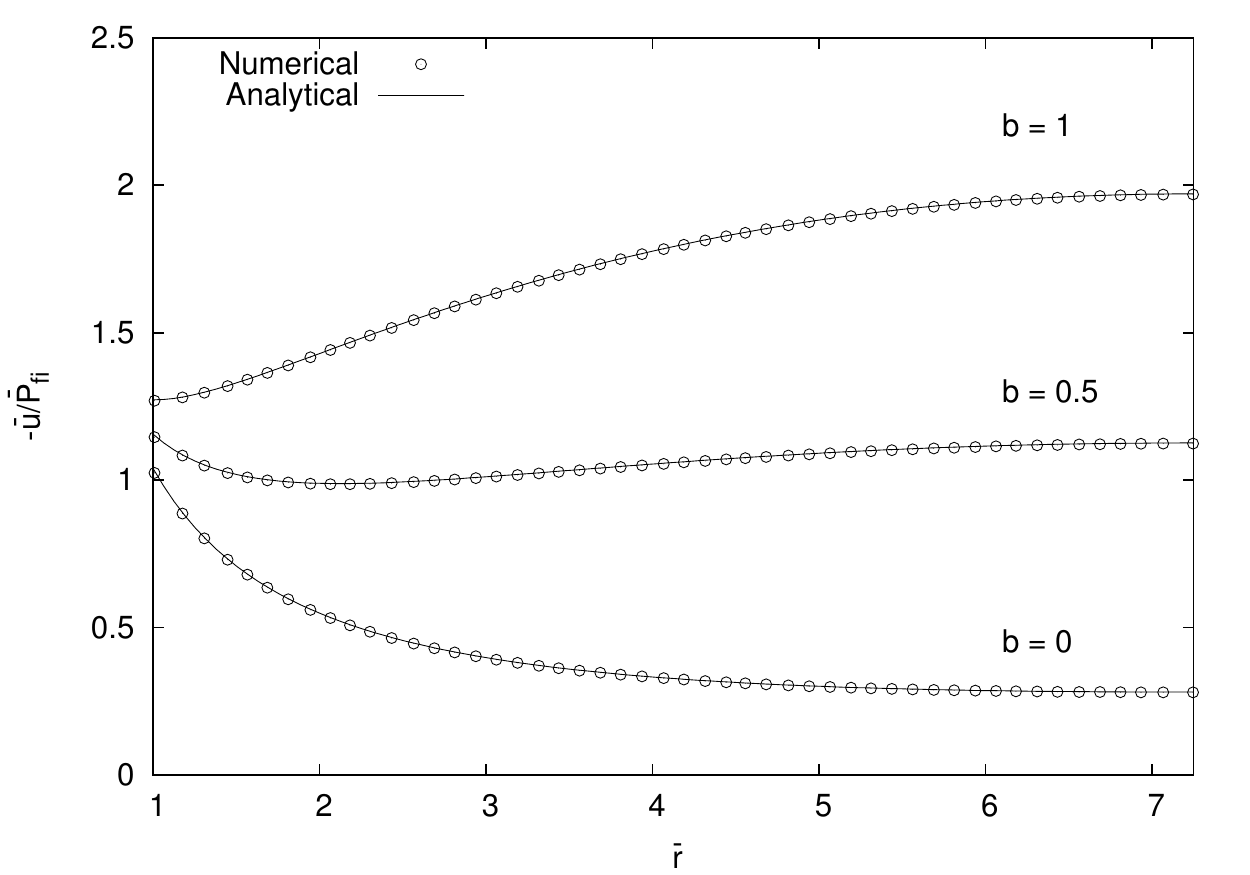}\\
(b) &  \includegraphics[width=10.cm]{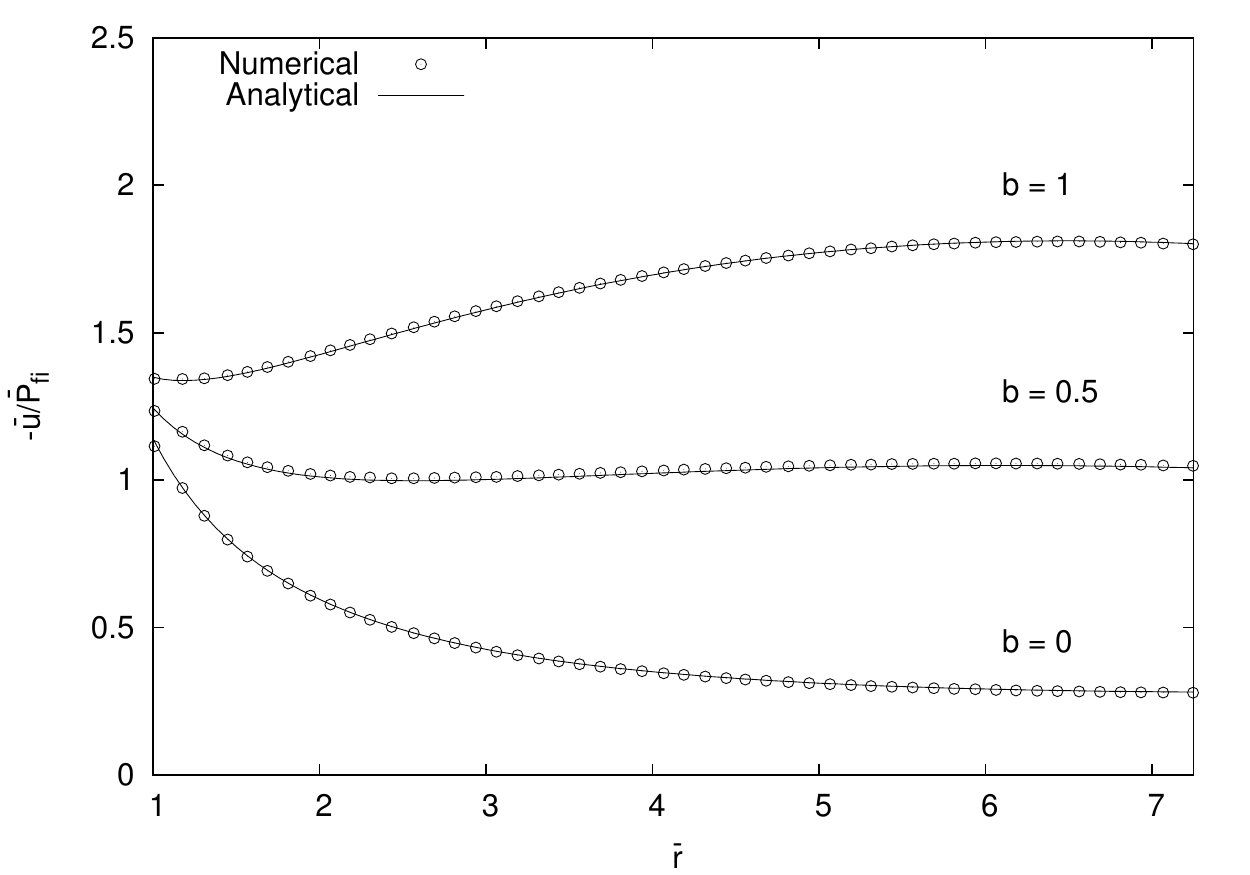}\\
(c) & \includegraphics[width=10.cm]{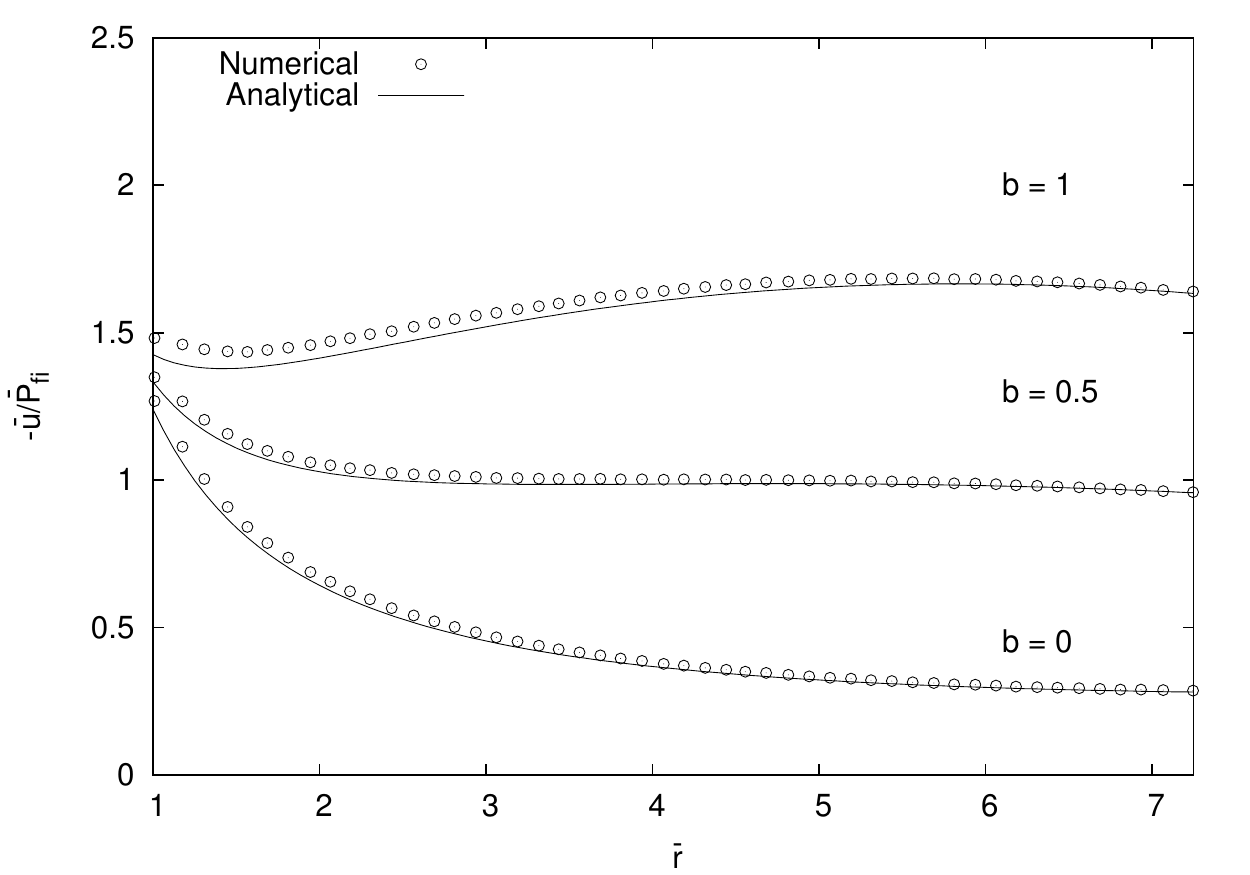}\\
\end{tabular}
\end{center}
 \caption{Comparison of numerical and analytical results in the form of normalised radial displacement versus normalised radial coordinate for Biot's coefficients $b=0$, $0.5$ and $1$, and Poisson's ratios (a) $\nu=0$, (b) $\nu=0.1$ and (c) $\nu=0.2$.}
\label{fig:couNu}
\end{figure}

In Figure~\ref{fig:couNu}, the results for the normalised radial displacements, $\bar{u} = u/r_{\rm i}$, obtained with the lattice models are compared with the analytical solution in (\ref{eq:anal21}) for varying Poisson's ratios $\nu=0$, $\nu=0.1$ and $\nu=0.2$.
For this range of Poisson's ratio, the present irregular lattice approach is expected to give acceptable results, as discussed in Section~\ref{subsec:MechLaw}.  
For each value of Poisson's ratio, the response for Biot's coefficients of $b = 0$,~$0.5$~and~$1$ are investigated.
The lattice model reproduces well the analytical solution for the radial displacements of the thick-walled cylinders.
In particular, for Poisson's ratio of $\nu=0$ and $0.1$, the agreement is very good.
Also, for $\nu=0.2$ a good agreement is obtained, but small deviations close to the inner boundary can be observed.
For all Poisson's ratios, the Biot's coefficient has a strong influence on the radial displacement. 
With a compressive internal fluid pressure ($P_{\rm fi}$), the wall thickness increases for $b = 1$, whereas it decreases for $b=0$.

In the present elastic analyses, boundary coupling Approach~1 is used to couple the mechanical and transport analysis. 
In the next section, the model is applied to analyses of the initiation and propagation of fracture. These analyses are carried out with a displacement driven control, which is denoted as boundary coupling Approach~2 in Section~\ref{sec:coupling}.
To demonstrate that this alternative approach also results in a good agreement with the analytical solution, an additional comparison with the analytical solution is performed.
The elastic stiffness obtained from the lattice analyses and the analytical solution are compared by setting in (\ref{eq:anal21}) $\dr=1$ and $\nu=0$ and solving for the pressure $\dPf$, which gives
\begin{equation}
\dPfi = \bar{C} \du = \dfrac{1}{\left[ b \left( \dfrac{\dro^2}{1-\dro^2} - \dfrac{1}{2 \ln \dro} \right)   + \left(1-b\right) \dfrac{1+\dro^2}{1-\dro^2} \right]} \du 
\end{equation}
Here, $\dro$ is $r_{\rm{o}}/r_{\rm{i}}$.
 
\begin{figure}
\begin{center}
  \includegraphics[width=12.cm]{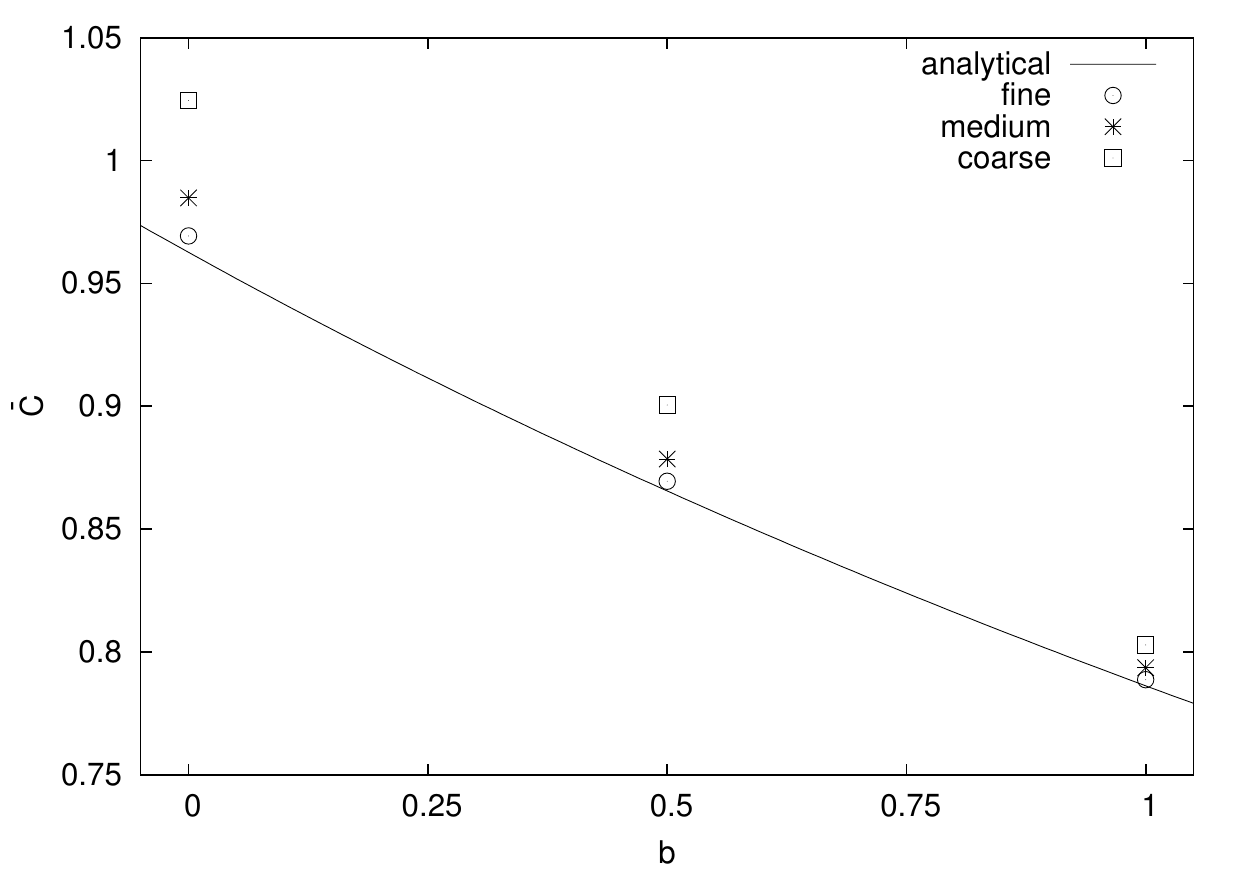}
\end{center}
\caption{Influence of Biot's coefficient on stiffness of thick-walled cylinder for boundary Approach~2.}
\label{fig:stiffness}
\end{figure}

The influence of Biot's coefficient $b$ on the normalised elastic stiffness $\bar{C}$ of the thick-walled cylinder is shown for both the lattice model and the analytical solution in Figure~\ref{fig:stiffness}.
For fine lattices the boundary Approach~2 reproduces well the stiffness obtained from the analytical solution. Coarse lattices overpredict the anlaytical stiffness, as expected for a displacement driven approach.

\section{Lattice analysis of hydraulic fracturing of a thick-walled cylinder}

\begin{figure}
\begin{center}
  \includegraphics[width=12.cm]{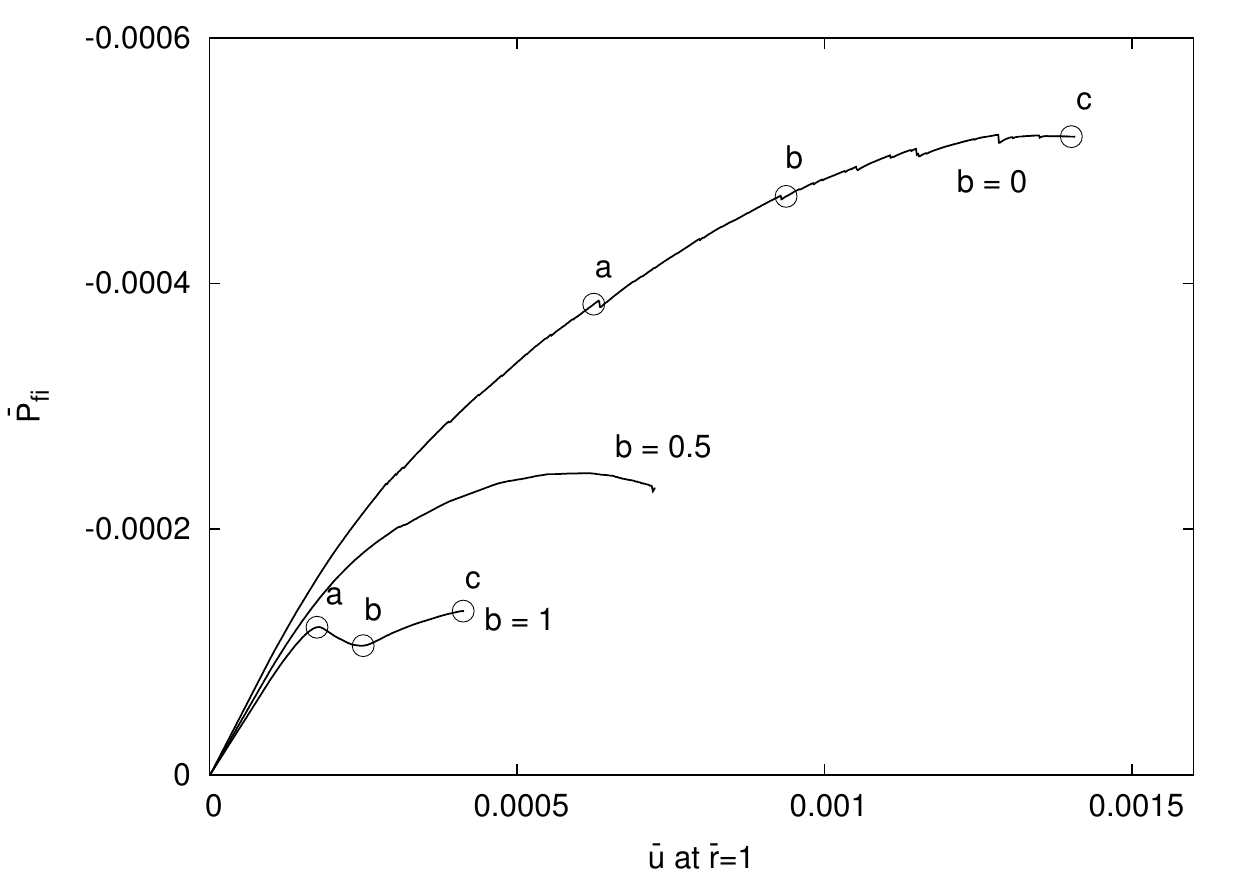}
\end{center}
\caption{Internal pressure versus radial displacement of inner boundary for $\nu=0$ and $b=0$, $0.5$ and $1$. The circles indicate stages at which the crack patterns are shown in Figures~\ref{fig:crackBiot0p0} and Figures~\ref{fig:crackBiot1p0} for $b=0$~and~$1$, respectively.}
\label{fig:compCrack}
\end{figure}

\begin{figure}
\begin{center}
\begin{tabular}{ccc}
  \includegraphics[width=4.5cm]{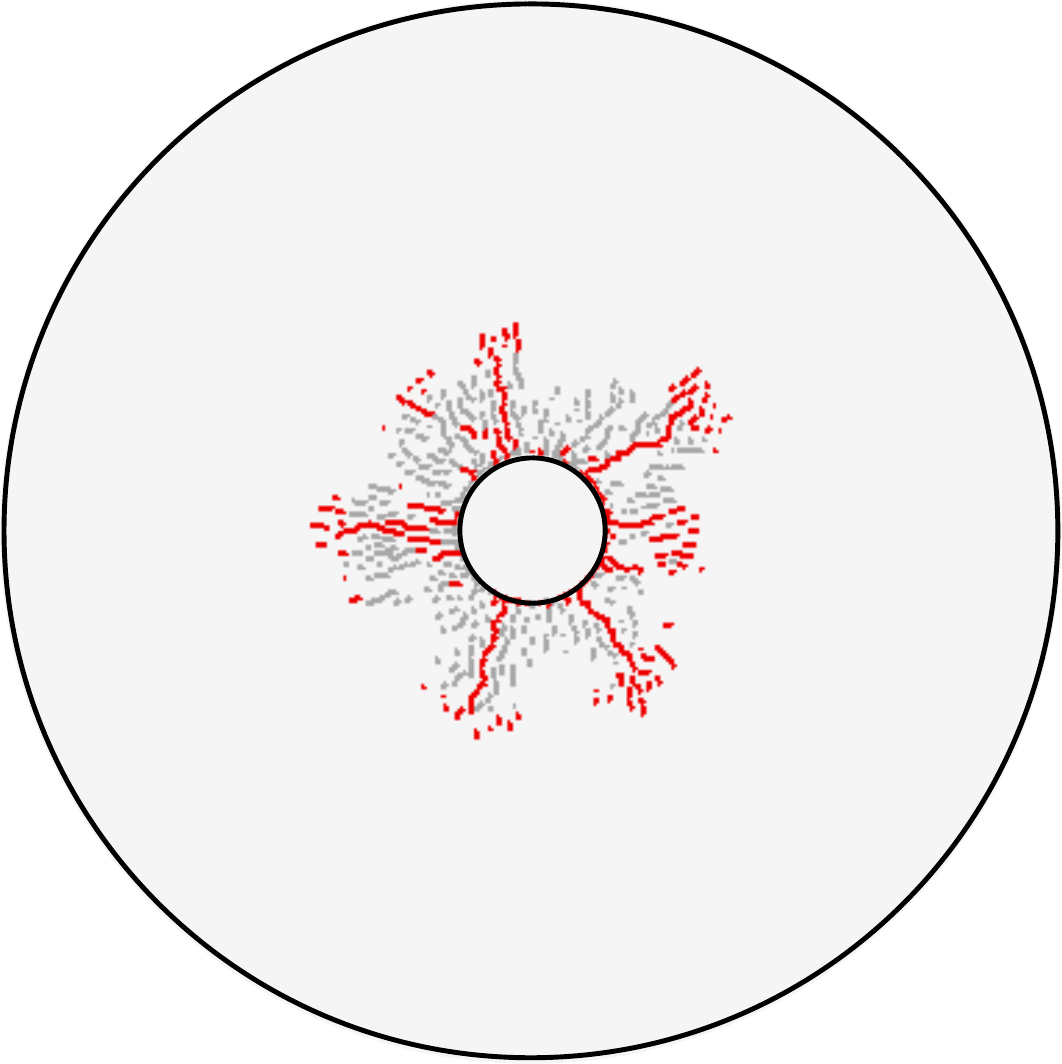} &   \includegraphics[width=4.5cm]{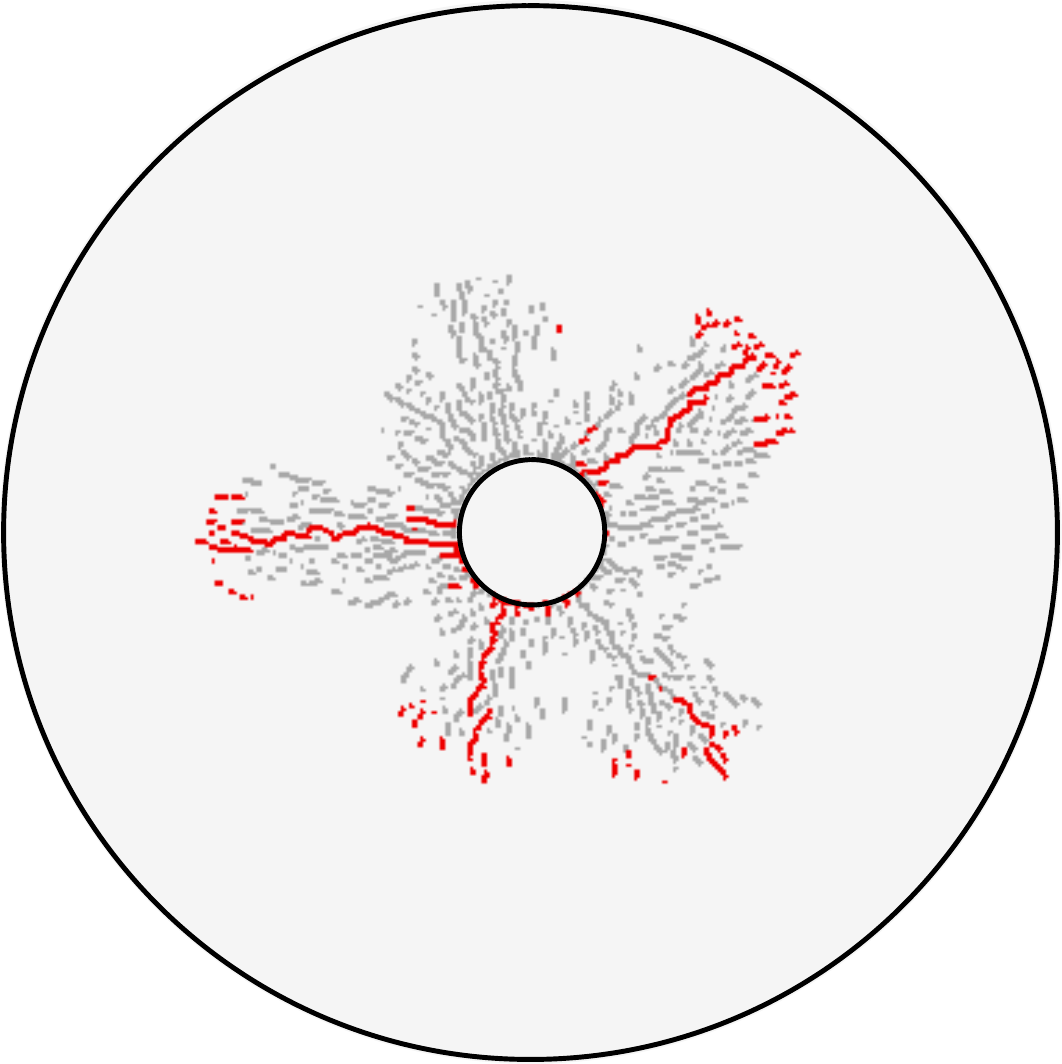}  &   \includegraphics[width=4.5cm]{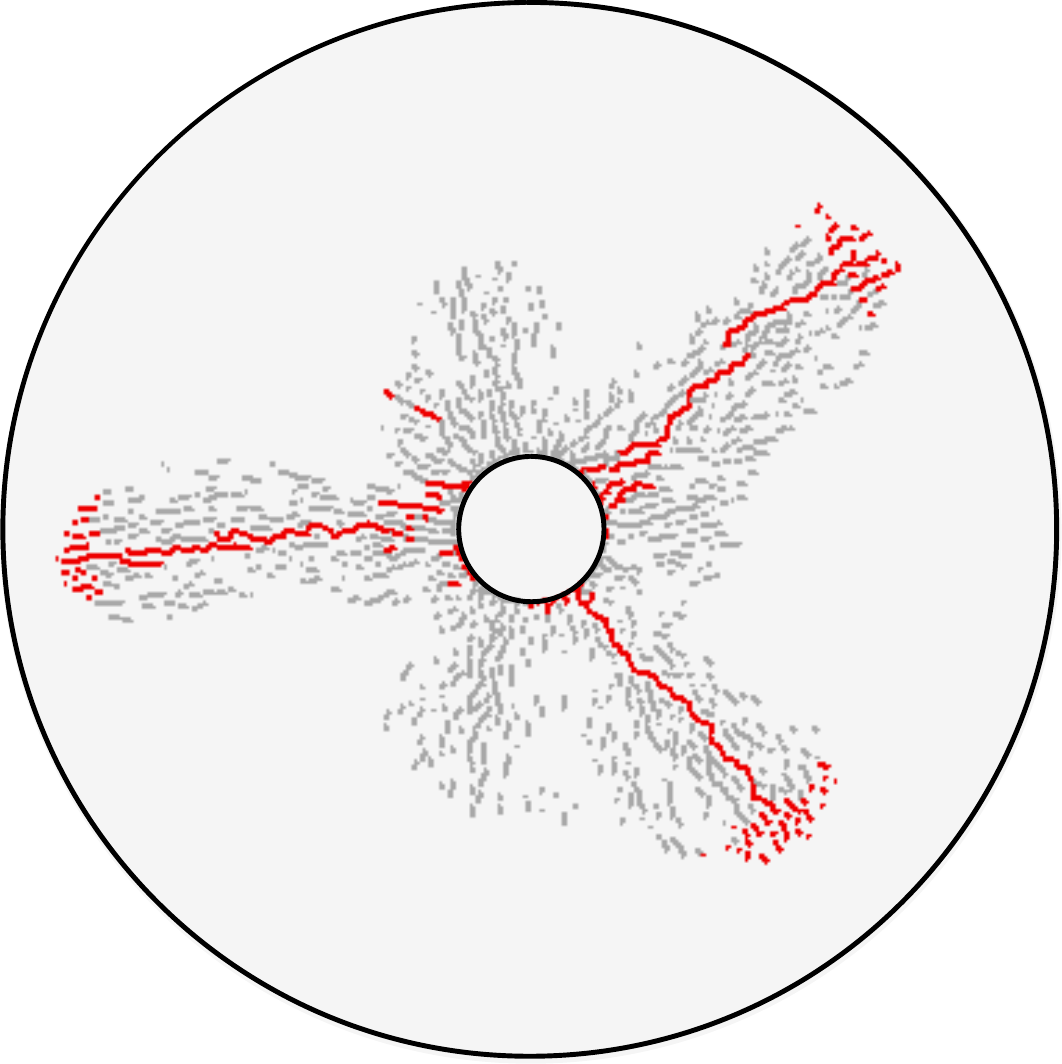}\\
(a) & (b) & (c)
\end{tabular}
\end{center}
 \caption{Crack patterns for $b=0$ at three stages shown in Figure~\ref{fig:compCrack}.}
\label{fig:crackBiot0p0}
\end{figure}

\begin{figure}
\begin{center}
\begin{tabular}{ccc}
 \includegraphics[width=4.5cm]{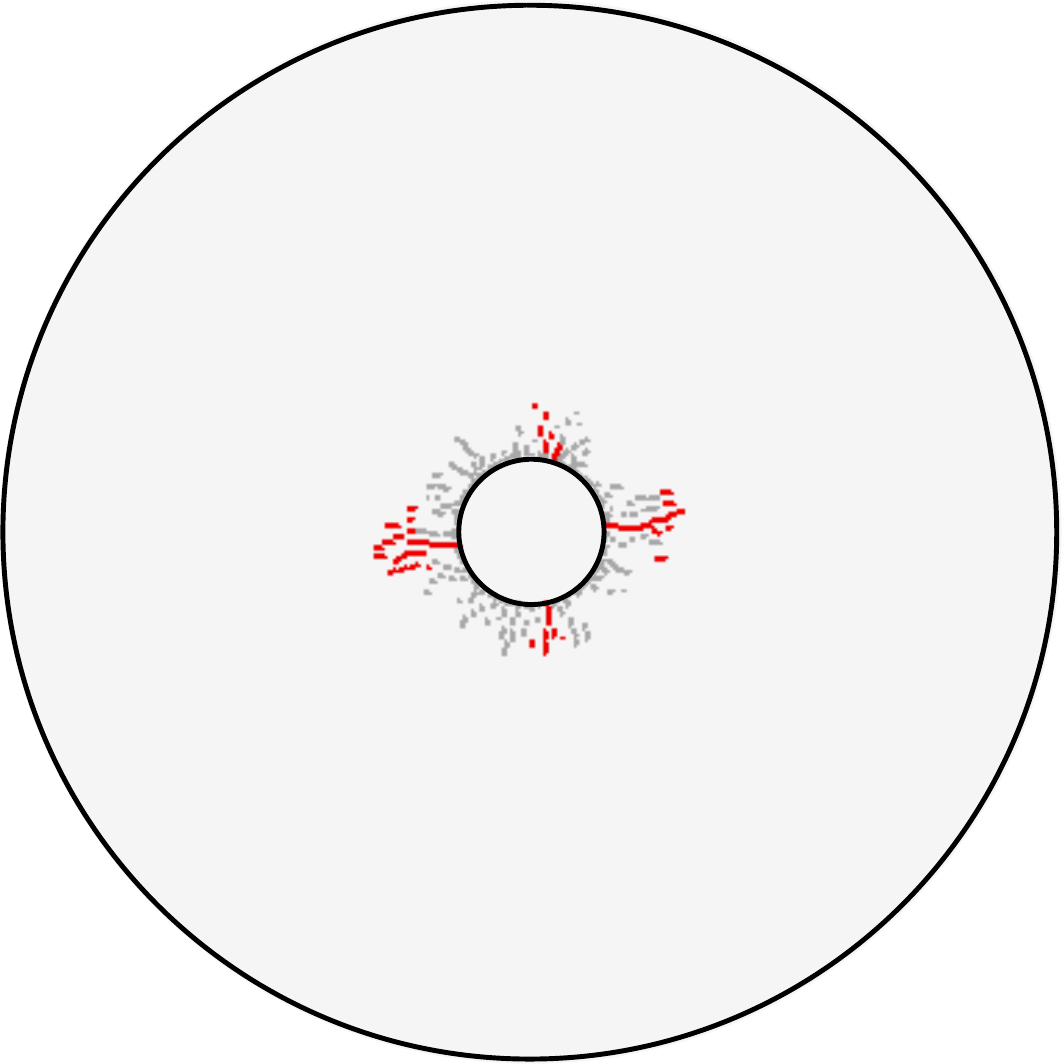} &   \includegraphics[width=4.5cm]{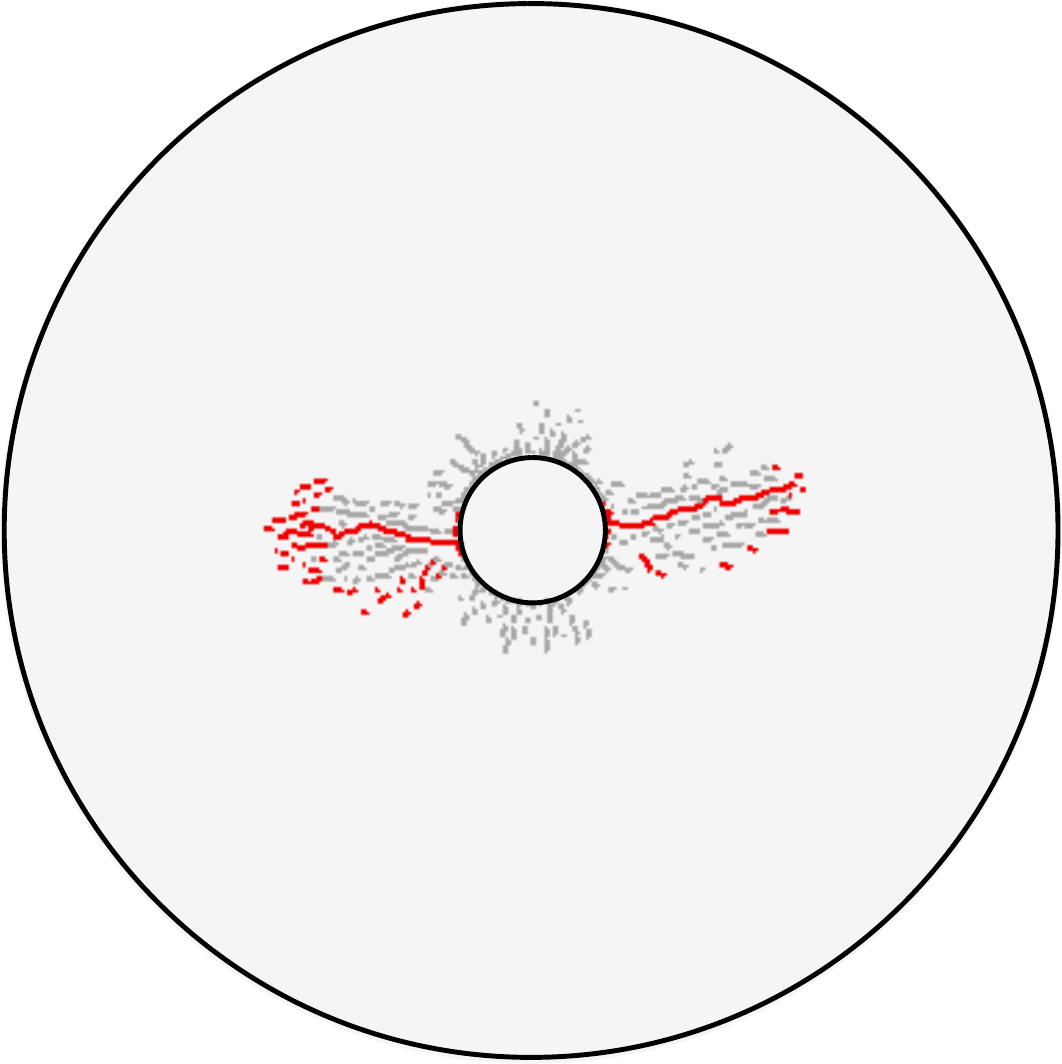}  &   \includegraphics[width=4.5cm]{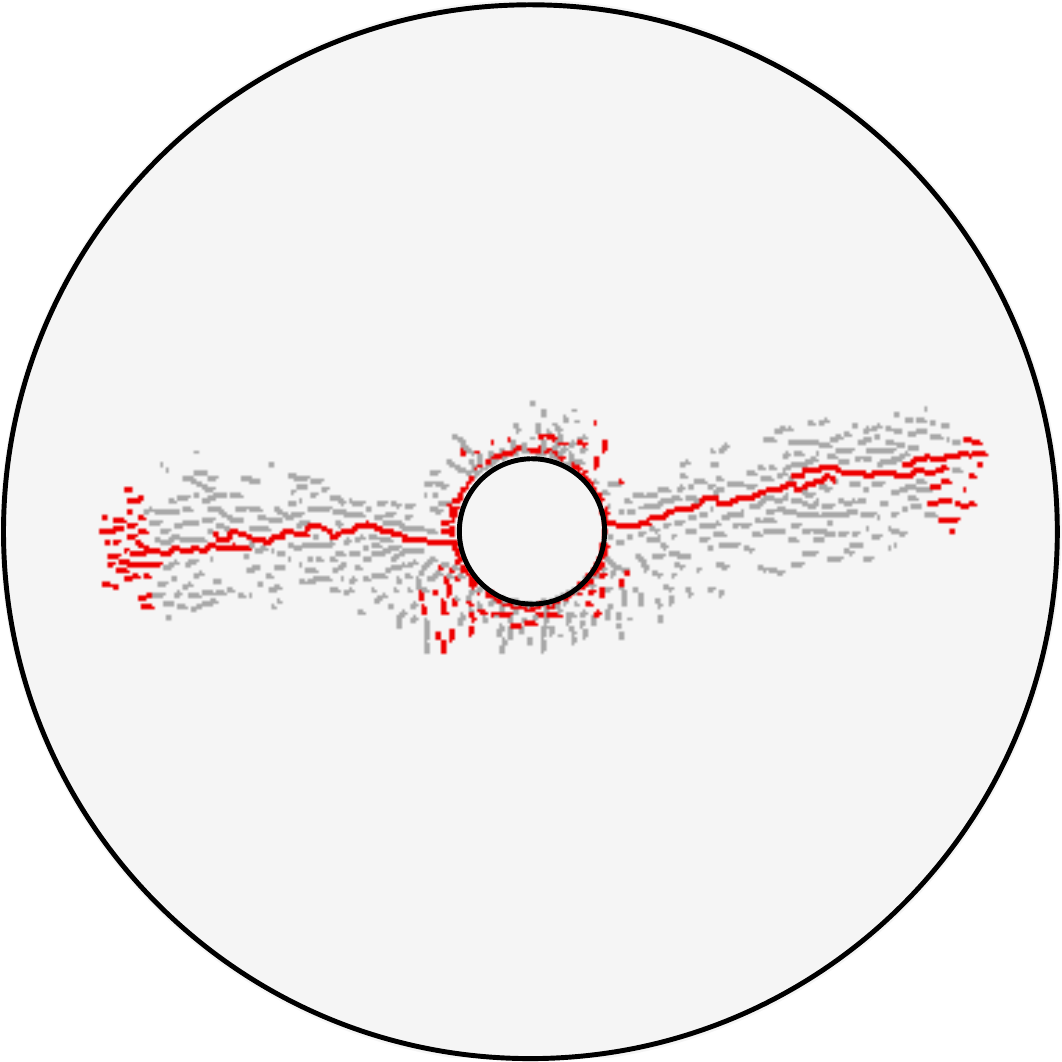}\\
(a) & (b) & (c)
\end{tabular}
\end{center}
 \caption{Crack patterns for $b=1$ at three stages shown in Figure~\ref{fig:compCrack}.}
\label{fig:crackBiot1p0}
\end{figure}

In this section, the influence of Biot's coefficient on crack initiation and propagation is investigated.
For these analyses, the coupling of boundary conditions follows approach 2. Thus, the analyses are controlled by the radial displacement at the inner boundary so that softening, defined here as a decrease of the fluid pressure on the inner boundary, $P_{\rm{fi}}$, with increasing radial displacement of the boundary, can be described.
The input parameters of the mechanical model are set to $\varepsilon_0 = 0.0001$, $\bar{w}_{\rm f} = w_{\rm f}/r_{\rm i} = 0.00625$, $q=2$ and $c =20$. For all fracture analyses, $\nu=0$ was assumed.
The lattice results are presented in Figure~\ref{fig:compCrack} in the form of normalised pressure versus normalised radial displacement at the inner boundary for $b=0$, $0.5$ and $1$.
The circles in Figure~\ref{fig:compCrack} indicate stages at which the crack patterns are shown in Figures~\ref{fig:crackBiot0p0}~and~\ref{fig:crackBiot1p0} for $b=0$ and $b=1$, respectively. 
For the crack patterns, dark grey lines (red in color) mark cross-sections of elements in which the damage parameter increases at this stage of analysis. Light grey lines represent cross-sections of elements which were damaged at a previous stage of analysis, but for which damage does not increase at this stage.  
Similar to the elastic analyses, Biot's coefficient has a strong influence on both the pressure-displacement curves in Figure~\ref{fig:compCrack} and the crack patterns in Figures~\ref{fig:crackBiot0p0}~and~\ref{fig:crackBiot1p0}.
With increasing Biot's coefficient, the peak load decreases significantly.
Also, the evolution of the crack patterns differs strongly. 
For $b=0$, three dominant cracks propagate outwards, whereas for $b=1$ only two dominant cracks are visible.

\begin{figure}
\begin{center}
  \includegraphics[width=12.cm]{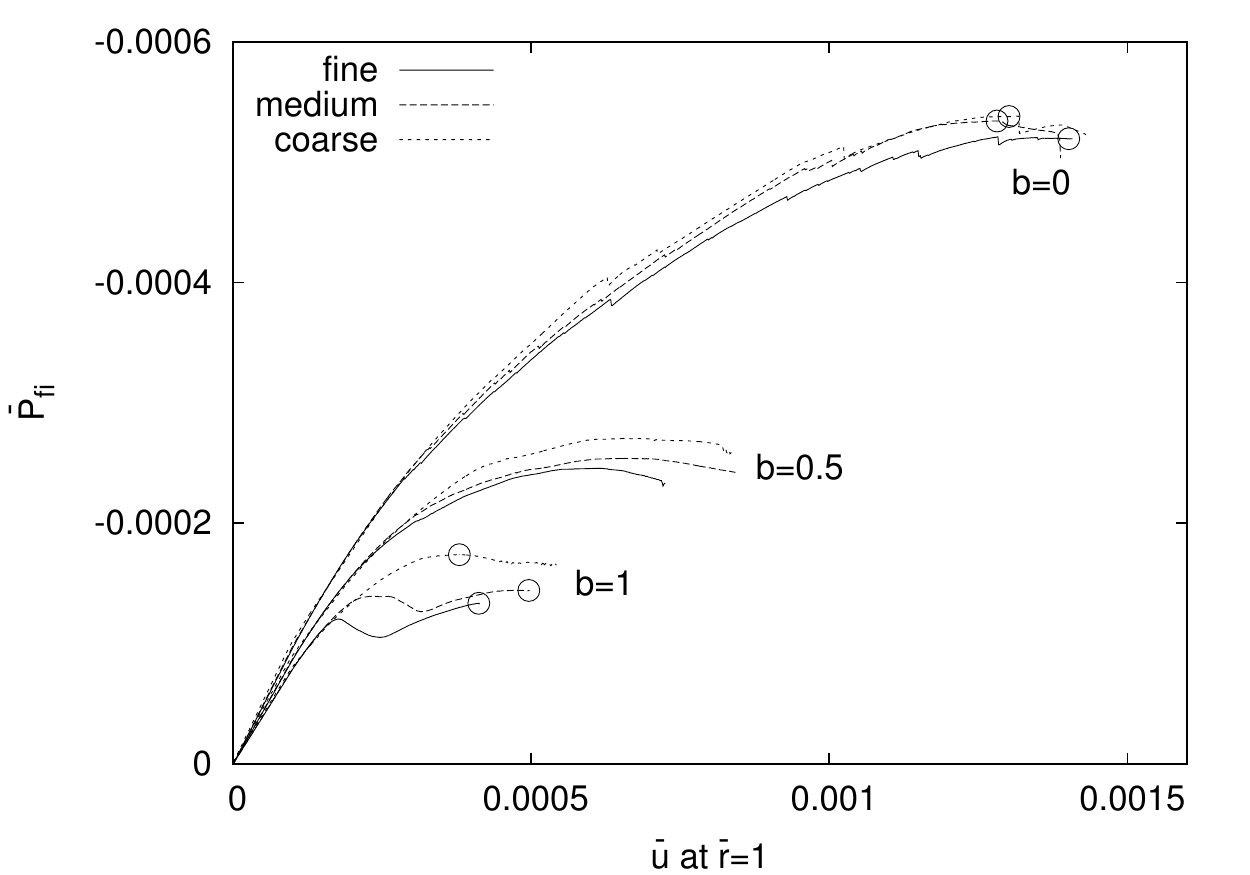}
\end{center}
\caption{Internal pressure versus radial displacement of inner boundary for three mesh-sizes and three Biot's coefficients. The circles indicate stages at which crack patterns are shown in Figures~\ref{fig:crackMeshBiot0p0}~and~\ref{fig:crackMeshBiot1p0} for $b=0$~and~$1$, respectively.}
\label{fig:meshCrack}
\end{figure}

\begin{figure}
\begin{center}
\begin{tabular}{ccc}
 \includegraphics[width=4.5cm]{./figCrackBiot0p0_1dmin.pdf} & \includegraphics[width=4.5cm]{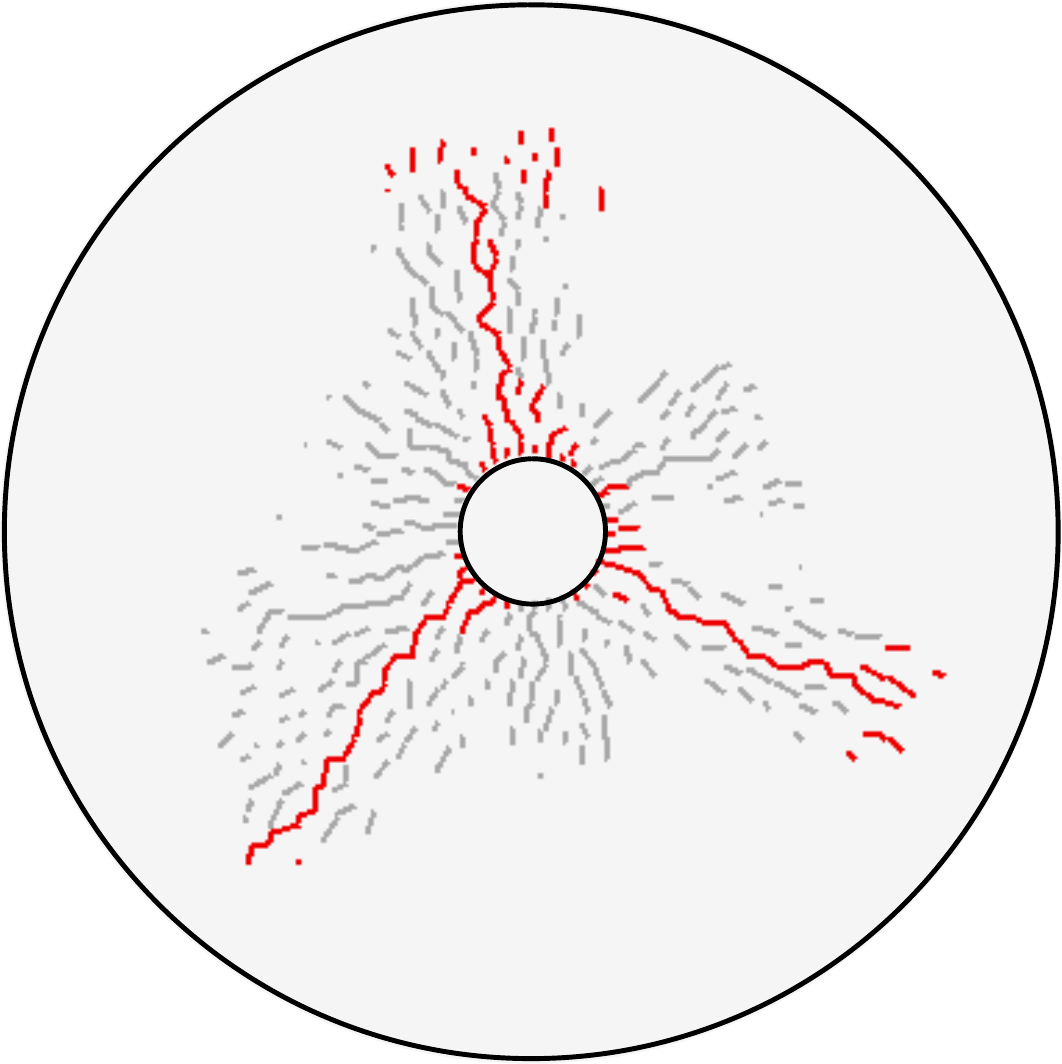}  &  \includegraphics[width=4.5cm]{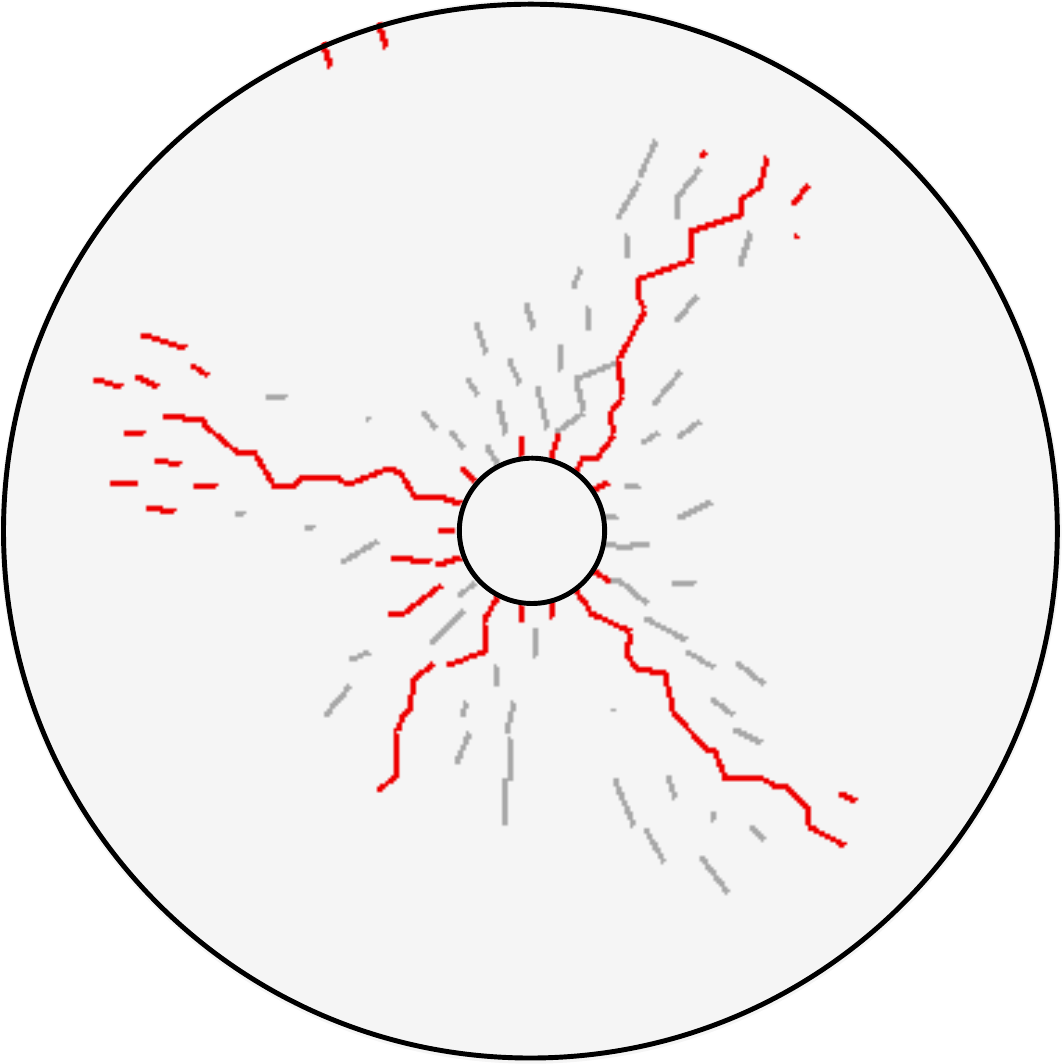}\\
(a) & (b) & (c)
\end{tabular}
\end{center}
 \caption{Crack patterns for 3 mesh-sizes for the analysis for $b = 0$ at peak for (a) fine, (b) medium and (c) coarse lattices.}
\label{fig:crackMeshBiot0p0}
\end{figure}

\begin{figure}
\begin{center}
\begin{tabular}{ccc}
\includegraphics[width=4.5cm]{./figCrackBiot1p0_1dmin.pdf} & \includegraphics[width=4.5cm]{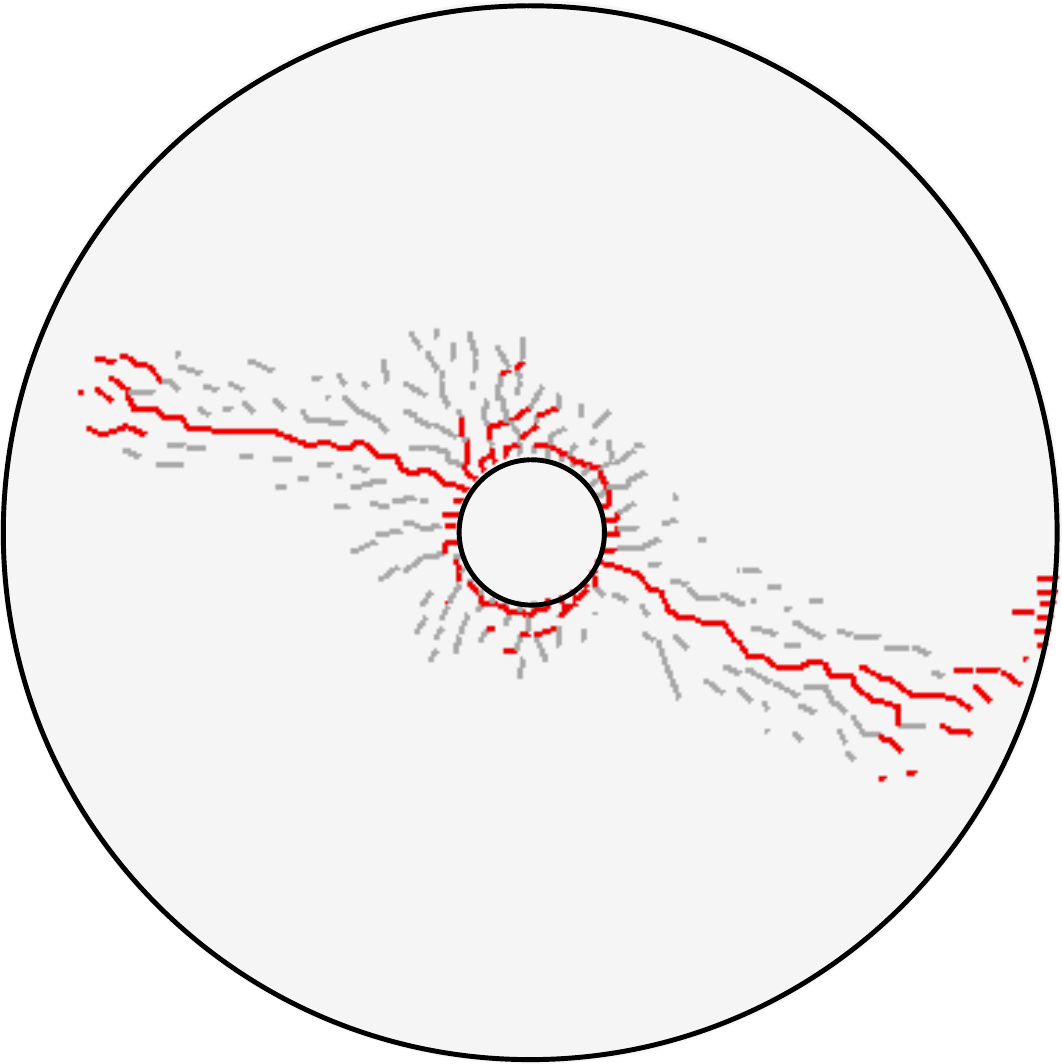} & \includegraphics[width=4.5cm]{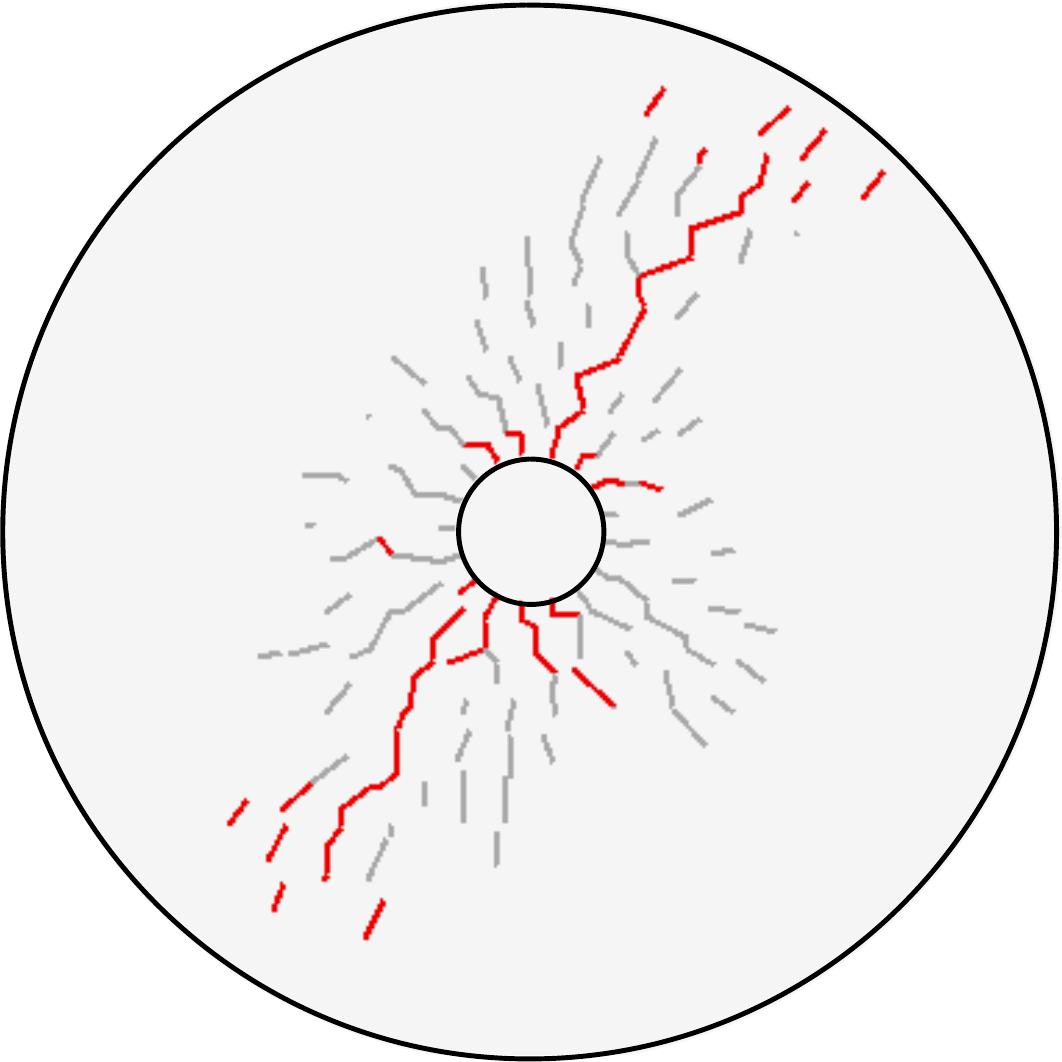}\\
(a) & (b) & (c)
\end{tabular}
\end{center}
 \caption{Crack patterns for 3 mesh-sizes for $b = 1$ at peak for (a) fine, (b) medium and (c) coarse lattices.}
\label{fig:crackMeshBiot1p0}
\end{figure}

The model performance for describing hydraulic fracturing is further investigated by studying mesh-size dependence, and agreement with upper and lower bounds. 
Concerning the mesh-size, its influence on pressure-displacement curves and crack patterns was investigated by comparing the results for $\bar{d}_{\rm{min}} = 0.123$, $0.246$ and $0.492$, which are denoted as fine, medium and coarse, respectively. 
 In Figure~\ref{fig:meshCrack}, the influence of the mesh-size on three pressure-displacement curves for the three Biot's coefficients are presented. 
Furthermore, the mesh-size influence on crack patterns at peak for $b=0$ and $b=1$ are shown in Figures~\ref{fig:crackMeshBiot0p0}~and~\ref{fig:crackMeshBiot1p0}.
Both pressure-displacement and crack patterns are almost independent of the mesh-size. Concerning the crack patterns, the orientation of cracks for the present axissymmetric problem is arbitrary and differ for the three mesh-sizes. However, the number of cracks and their length at peak are almost independent of the mesh-size. 

\begin{figure}
\begin{center}
\includegraphics[width=5cm]{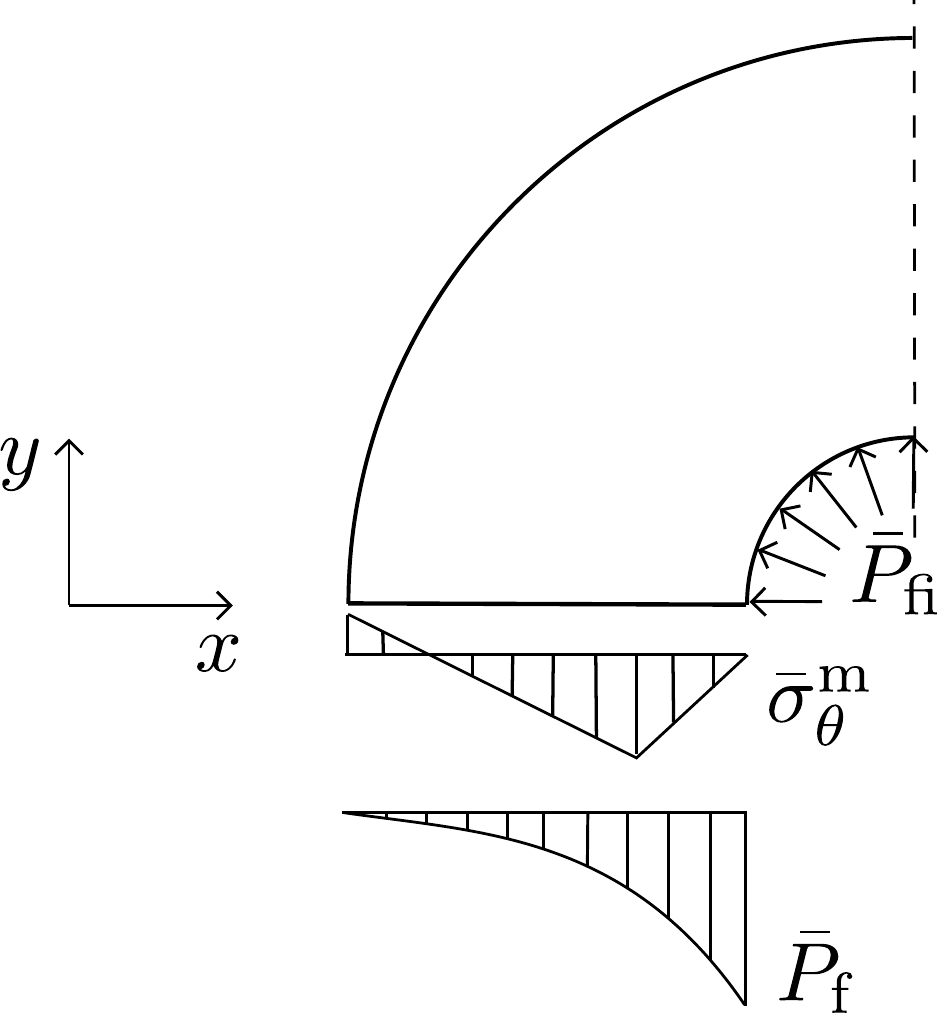}
\end{center}
 \caption{Schematic force equilibrium with the total circumferential stress $\bar{\sigma}_{\rm \theta}$ composed of a mechanical and a fluid part.}
\label{fig:plasticLimit}
\end{figure}

The influence of Biot's coefficient $b$ on the maximum load capacity for the lattice model is further assessed by comparing it with lower and upper limits obtained from the analytical solution.
Force equilibrium in the y-direction, as shown schematically in Figure~\ref{fig:plasticLimit}, gives
\begin{equation}\label{eq:integral}
\bar{P}_{\rm fi} = - \int_1^{\dro} \left(\bar{\sigma}_{\rm \theta}^{\rm m} + \dPf \right) d\bar{r}    
\end{equation}
where $\dPf$ is the fluid pressure in (\ref{eq:dimFlow}) and $\bar{\sigma}_{\rm \theta}^{\rm m}$ is the mechanical circumferential stress, which,in general, are both functions of the radial coordinate $\bar{r}$.
For the upper limit, the mechanical circumferential stress is assumed to be constant in the radial direction with a value $\bar{\sigma}_{\rm \theta}^{\rm m} = f_{\rm t}/E = \varepsilon_0$. For this case, inserting $\bar{P}_{\rm f}$ from (\ref{eq:dimFlow}) into (\ref{eq:integral}) results in
\begin{equation}\label{eq:upperLimit}
\bar{P}_{\rm fi} = - \dfrac{(\dro - 1)\varepsilon_{\rm 0}}{1 - b \left(\dfrac{1}{\ln \dro} + \dfrac{1}{(\ln \dro)^2}-\dfrac{\dro}{\ln \dro}\right)}
\end{equation}

The lower limit is chosen here as the inner pressure at which cracking initiates.
It is derived by setting $\bar{\sigma}_{\rm \theta}^{\rm m}$ in (\ref{eq:anal23}) equal to $\varepsilon_0$ for $\nu=0$ and $\dr = 1$ and solving for the pressure, which gives  
\begin{equation}\label{eq:lowerLimit}
\bar{P}_{\rm fi} = - \dfrac{\varepsilon_0} {\left[ b \left(\dfrac{\dro^2}{\dro^2-1} + \dfrac{1}{2 \ln \dro}\right) + (1-b)\dfrac{1+\dro^2}{\dro^2-1} \right]}
\end{equation}

\begin{figure}
\begin{center}
  \includegraphics[width=12.cm]{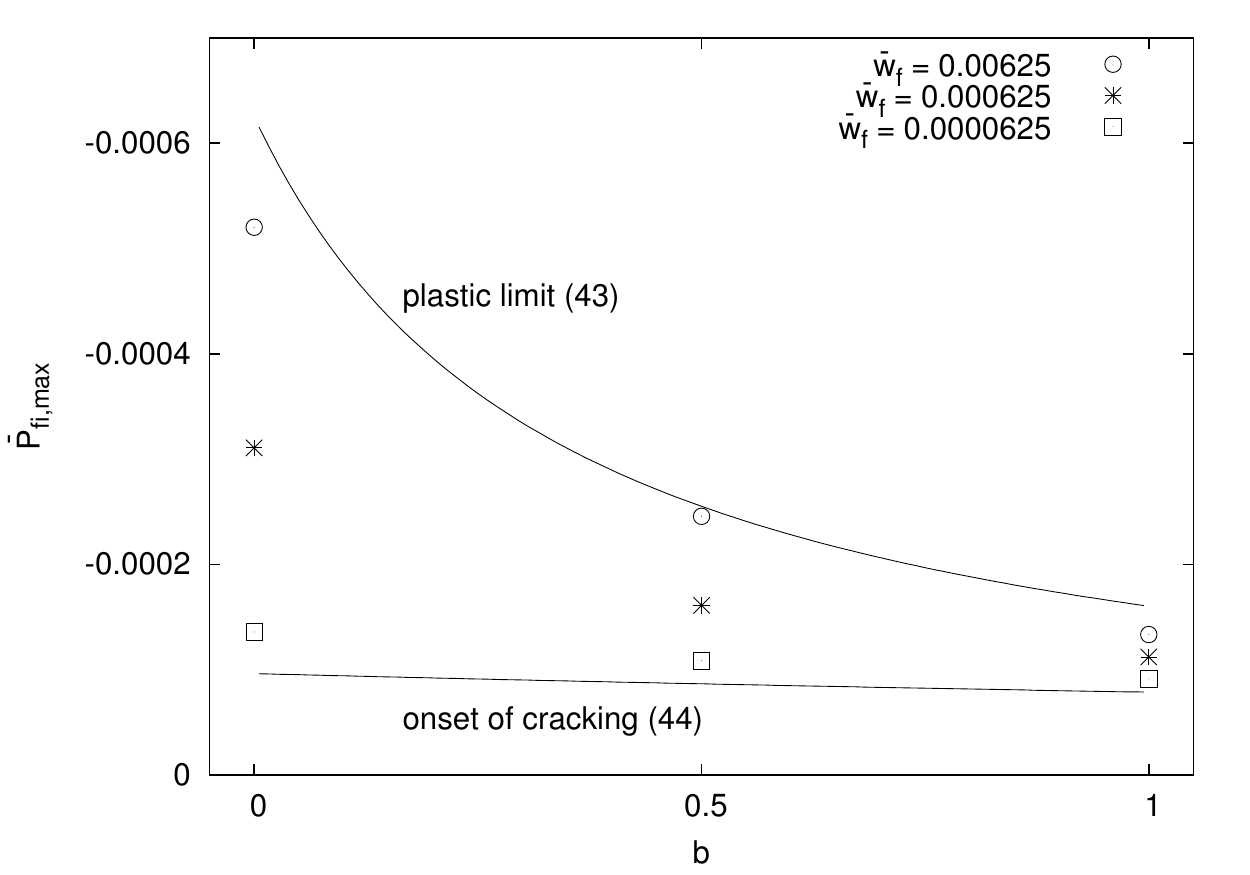}
\end{center}
\caption{Comparison of peak loads of lattice analyses with analytical limits for varying Biot coefficients.}
\label{fig:limits}
\end{figure}

Three values of $\bar{w}_{\rm f}$ were chosen as $0.00625$,~$0.000625$~and~$0.0000625$. The comparison of peak loads obtained from lattice analyses with different $\bar{w}_{\rm f}$ and the upper and lower limits in (\ref{eq:upperLimit}) and (\ref{eq:lowerLimit}) is shown in Figure~\ref{fig:limits}.
For $\bar{w}_{\rm f} = 0.00625$, the peak load obtained from the lattice analyses is in good agreement with the analytical plastic limit. This indicates that this value of $\bar{w}_{\rm f}$ is so large with respect to the thickness of the cylinder ($\dro-1$), that the response at peak is close to the theoretical plastic limit. A smaller value of $\bar{w}_{\rm f}$ results in a reduction of the maximum load obtained in the lattice analyses. All values are positioned between the onset of cracking and the upper plastic limit.
Since $\bar{w}_{\rm f} = w_{\rm f}/r_{\rm i}$ is a dimensionless quantity, which depends on both $w_{\rm f}$ and $r_{\rm i}$, the influence of it on the normalised load capacity of the thick-walled cylinder can be interpreted in two ways.
On the one hand, an increase of $\bar{w}_{\rm f}$ can be interpreted as an increase of $w_{\rm f}$ (increase of fracture energy) at a constant inner radius $r_{\rm i}$. For this case, the results in Figure~\ref{fig:limits} show that an increase of the fracture energy results in an increase of the normalised load capacity.
On the other hand, a decrease of the size of the cylinder, represented by $r_{\rm i}$, at a constant fracture energy, represented by $w_{\rm f}$, also results in an increase of $\bar{w}_{\rm f}$. Therefore, the smaller the cylinder is at constant fracture energy, the greater is the normalised load capacity. This type of size effect is well known for structures made of quasi-brittle materials, which undergo softening \citep{Baz01}. For increasing Biot's coefficient, this size effect decreases significantly, because the fluid pressure is independent of the specimen size, but the mechanical stress distribution is size dependent.

\section{Conclusions}
A hydro-mechanical lattice approach for modelling hydraulic fracture has been proposed and applied to a thick-walled cylinder subjected to an internal pressure. 
The lattice results were compared to the analytical solution for the elastic case derived in the present study.
The elastic response obtained with the lattice model agrees very well with the analytical solution for varying Biot's coefficient and Poisson's ratio.
With high values of Biot's coefficient, the wall thickness of the cylinder increases. With low values of Biot's coefficient, the thickness decreases.
The fracture analyses show that the lattice approach can describe hydraulic cracking mesh-size independently.
With increasing Biot's coefficient, the peak pressure decreases strongly. Furthermore, the thick walled-cylinder exhibits a strong size effect on the nominal strength. This size effect decreases with an increase in Biot's coefficent.

\section*{Acknowledgements}
The numerical analyses were performed with the nonlinear analyses program OOFEM \citep{Pat12} extended by the present authors.
The first and fourth authors acknowledge funding received from the UK Engineering and Physical Sciences Research Council (EPSRC) under grant EP/I036427/1.

\bibliographystyle{elsarticle-harv}
\bibliography{general}

\appendix
\numberwithin{equation}{section}
\section{Appendix} \label{sec:appendix}
In the present section, the analytical solution of the fluid pressure distribution and the elastic response of a thick-walled cylinder subjected to internal fluid pressure under steady-state conditions is presented (Figure~\ref{fig:cylinder}).
Firstly, we determine the fluid pressure distribution across the cylinder.
The radial component $q$ of the flux vector $\mathbf{q}$ at a give distance $r$ from the centre of the thick-walled cylinder is calculated by imposing conservation of fluid mass as 
\begin{equation} \label{eq:anal1}
q = \dfrac{Q}{2\pi r t}
\end{equation}
where $Q$ is the total flow rate and $t$ is the out of plane cylinder thickness.
Due to axissymmetry, the tangential component of flux is equal to zero.
The radial component of the flux vector can be alternatively calculated from Darcy's law as
\begin{equation}\label{eq:anal2}
q = k \dfrac{dP_{\rm f}}{d r}
\end{equation}  
As stated earlier, the sign convention for pore fluid pressure $P_{\rm f}$ is tension positive.

Setting the right hand sides of (\ref{eq:anal1}) and (\ref{eq:anal2}) equal and then integrating,
\begin{equation}\label{eq:anal3}
P_{\rm f} = \dfrac{\mu Q}{2 \kappa \pi t} \ln r  + C
\end{equation}
is obtained.
Here, $C$ is a constant of integration, which is determined as
\begin{equation}\label{eq:anal4}
C = -\dfrac{Q}{2 k \pi t} \ln r_o
\end{equation}
by imposing the boundary condition of fluid pressure at the outer boundary $P_{\rm f}(r = r_{\rm o}) = 0$.

Setting (\ref{eq:anal4}) in (\ref{eq:anal3}) gives
\begin{equation} \label{eq:anal5}
P_{\rm f} = -\dfrac{Q}{2 k \pi t} \ln \dfrac{r_o}{r}
\end{equation}

By imposing the boundary condition of fluid pressure at the inner boundary ($P_{\rm f}(r = r_{\rm i}) = P_{\rm fi}$),  the following expression for $Q$ is obtained from (\ref{eq:anal5}):
\begin{equation} \label{eq:anal6}
Q = - \dfrac{2 k \pi t}{\ln \dfrac{r_{\rm o}}{r_{\rm i}}}P_{\rm fi}
\end{equation}
By setting (\ref{eq:anal6}) into (\ref{eq:anal5}), an alternative expression of the fluid pressure $P_{\rm f}$ is obtained:
\begin{equation}\label{eq:anal7}
P_{\rm f} =  P_{\rm fi} \dfrac{\ln \dfrac{r_{\rm o}}{r}}{\ln \dfrac{r_{\rm o}}{r_{\rm i}}}
\end{equation}
which is a function of $r$, $r_{\rm i}$, $r_{\rm o}$ and $P_{\rm fi}$, but independent of the hydraulic conductivity $k$. 
Introducing the dimensionless variables, $\bar{r} = r/r_{\rm i}$, $\bar{r}_{\rm o} = r_{\rm o}/r_{\rm i}$, $\dPf = P_{\rm f}/E_{\rm c}$ and $\dPfi = P_{\rm fi}/E_{\rm c}$ (\ref{eq:anal7}) is written in dimensionless form as
\begin{equation} \label{eq:dimFlow}
\dPf = \dPfi \dfrac{\ln \dfrac{\dro}{\dr}}{\ln \dro}
\end{equation}

Secondly, we calculate the radial displacement caused by the fluid pressure in the cylinder.
We start from the equilibrium equation of the thick-walled cylinder, which, under axissymmetric conditions, is given for instance in \citet{TimGoo87} as
\begin{equation}\label{eq:anal8}
\dfrac{d\sigma_{\rm r}}{dr} + \dfrac{\sigma_{\rm r} - \sigma_{\rm \theta}}{r} = 0
\end{equation}
where $\sigma_{\rm r}$ and $\sigma_{\rm \theta}$ are the total radial and tangential stress, respectively, that are equal to the sum of effective (mechanical) stress and the contribution of the pore fluid pressure, i.e. $\sigma_{\rm r} = \sigma_{\rm r}^{\rm m} + b P_{\rm f}$ and $\sigma_{\rm \theta} = \sigma_{\rm \theta}^{m} + b P_{\rm f}$. 
Here, $b$ is Biot's coefficient introduced previously in (\ref{eq:StressStrain}).

By substituting these total stresses in (\ref{eq:anal8}), one obtains the equilibrium equation expressed in terms of effective stresses and fluid pressure:
\begin{equation} \label{eq:anal9}
\dfrac{d \sigma_{\rm r}^{\rm m}}{dr} + \dfrac{\sigma_{\rm r}^{\rm m} - \sigma_{\rm \theta}^{\rm m}}{r} +  b \dfrac{d P_{\rm f}}{dr} = 0 
\end{equation}
This equilibrium equation is now further manipulated to obtain the displacement and stress field for the thick-walled cylinder.
Firstly, the expression of $P_{\rm f}$ in (\ref{eq:anal7}) is set into (\ref{eq:anal9}).
Then, the effective radial and tangential stresses are related to the corresponding strains by Hooke's law for plane stress conditions as
\begin{equation}  \label{eq:anal10}
  \sigma_{\rm r}^{\rm m} = \dfrac{E_{\rm c}}{1-\nu^2} \left(\varepsilon_{\rm r} + \nu \varepsilon_{\rm \theta}\right)
\end{equation}
and 
\begin{equation} \label{eq:anal11}
\sigma_{\rm \theta}^{\rm m} = \dfrac{E_{\rm c}}{1-\nu^2}\left(\varepsilon_{\rm \theta} + \nu \varepsilon_{\rm r}\right)
\end{equation}
Finally, kinematic relations are used to relate the radial and tangential strains, $\varepsilon_{\rm r}$ and $\varepsilon_{\rm \theta}$, respectively, to the radial displacement $u$, i.e. $\varepsilon_{\rm r} = \dfrac{du}{dr}$ and $\varepsilon_{\rm \theta} = \dfrac{u}{r}$.
These steps result in a differential equation for the radial displacement $u$ of the form
\begin{equation} \label{eq:anal12}
\dfrac{d^2u}{dr^2} + \dfrac{du}{dr}\dfrac{1}{r}-\dfrac{u}{r^2} - b \dfrac{P_{\rm fi}}{E_{\rm c}} \dfrac{1-\nu^2}{\ln \dfrac{r_{\rm o}}{r_{\rm i}}}\dfrac{1}{r} = 0
\end{equation}
This differential equation is now written in dimensionless variables $\du = u/r_i$, $\dro = r_{\rm o}/r_{\rm i}$, $\dr = r/r_{\rm i}$, $\dPfi = P_{\rm fi}/E_{\rm c}$ as 
\begin{equation} \label{eq:anal13}
\dfrac{d^2\du}{d\dr^2} + \dfrac{d\du}{d\dr}\dfrac{1}{\dr}-\dfrac{\du}{\dr^2} - b \dPfi \dfrac{1-\nu^2}{\ln \dro}\dfrac{1}{\dr} = 0
\end{equation}
The solution to (\ref{eq:anal13}) is
\begin{equation}\label{eq:anal14}
\du = \dfrac{1}{2} b \dPfi \dfrac{1-\nu^2}{\ln \dro} \dr \ln{\dr} + \dfrac{C_1}{\dr} + C_2 \dr
\end{equation}
with two integration constants $C_1$ and $C_2$.
This solution is used to determine the strains and stresses.
The radial and tangential strains are calculated from (\ref{eq:anal14}) as
\begin{equation} \label{eq:anal15}
\varepsilon_{\rm r} = \dfrac{d \du}{d \dr} =  \dfrac{1}{2} b \dPfi \dfrac{1-\nu^2}{\ln \dro} \left(\ln \dr + 1\right) - \dfrac{C_{1}}{\dr^2} + C_{2}
\end{equation}
and
\begin{equation} \label{eq:anal16}
\varepsilon_{\rm \theta} = \dfrac{u}{r} = \dfrac{1}{2} b \bar{P}_{\rm fi} \dfrac{1-\nu^2}{\ln \dro}\ln{\dr} + \dfrac{C_1}{\dr^2} + C_2
\end{equation}
respectively.

Similarly, the radial and tangential dimensionless stresses $\bar{\sigma}_{\rm r}^{\rm m} = \sigma_{\rm r}^{\rm m}/E_{\rm c}$ and $\bar{\sigma}_{\rm \theta}^{\rm m} = \sigma_{\rm \theta}^{\rm m}/E_{\rm c}$, respectively, are calculated from (\ref{eq:anal10}) and (\ref{eq:anal11}) and the above strain definitions in (\ref{eq:anal15}) and (\ref{eq:anal16}) as
\begin{equation} \label{eq:anal17}
\bar{\sigma}_{\rm r}^{\rm m} = \dfrac{\varepsilon_{\rm r} + \nu \varepsilon_{\rm \theta}}{1-\nu^2} = \dfrac{1}{2} b \dPfi \dfrac{1}{\ln \dro} \left[(1+\nu) \ln \dr + 1\right] - \dfrac{1}{1+\nu} \dfrac{C_1}{\dr^2} + \dfrac{1}{1-\nu}C_2
\end{equation}
and
\begin{equation} \label{eq:anal18}
\bar{\sigma}_{\rm \theta}^{\rm m} = \dfrac{\varepsilon_{\rm \theta} + \nu \varepsilon_{\rm r}}{1-\nu^2} = \dfrac{1}{2} b \dPfi \dfrac{1}{\ln \dro} \left[(1+\nu) \ln \dr + \nu \right] + \dfrac{1}{1+\nu} \dfrac{C_1}{\dr^2} + \dfrac{1}{1-\nu}C_2
\end{equation}

The radial stress component in (\ref{eq:anal17}) can now be used to determine the integration constants $C_{1}$ and $C_{2}$ in (\ref{eq:anal14}) by imposing that $\bar{\sigma}_{\rm r}^{\rm m} = (1-b) \dPfi$ at $\dr = 1$ and $\bar{\sigma}_{\rm r}^{\rm m} = 0$ at $\dr = \dro$. 
This yields
\begin{equation} \label{eq:anal19}
C_1 = - \dfrac{1}{2} b \dPfi \dfrac{\dro^2}{\dro^2-1} (1+\nu)^2 - (1-b) \dPfi \dfrac{\dro^2}{\dro^2-1} (1+\nu)
\end{equation}

\begin{equation} \label{eq:anal20}
C_2 = - \dfrac{1}{2} b \dPfi \left[\dfrac{\dro^2}{\dro^2-1}\left(1-\nu^2\right) + \dfrac{1-\nu}{\ln \dro}\right] - (1-b) \dPfi \dfrac{1-\nu}{\dro^2-1}
\end{equation}

Substituting (\ref{eq:anal19}) and (\ref{eq:anal20}) into (\ref{eq:anal14}), the following expression of the dimensionless displacement $\du$ is obtained:
\begin{equation}\label{eq:anal21}
\du = - b \dPfi \dfrac{1-\nu^2}{2} \left[\dfrac{\dro^2}{\dro^2-1} \left(\dfrac{1+\nu}{1-\nu}\dfrac{1}{\dr} + \dr\right) + \dr \dfrac{\dfrac{1}{1+\nu} - \ln \dr}{\ln \dro}\right] - (1-b) \dPfi \dfrac{\dro^2}{\dro^2-1} \left(\dfrac{1+\nu}{\dr} + \dfrac{\dr(1-\nu)}{\dro^2}\right)
\end{equation}
 
Furthermore, substituting (\ref{eq:anal19}) and (\ref{eq:anal20}) into equations (\ref{eq:anal17}) and (\ref{eq:anal18}) results in expressions for dimensionless radial and tangential stresses:
\begin{equation}\label{eq:anal22}
\bar{\sigma}_{\rm r}^{\rm m} = -\dfrac{1}{2} b \dPfi \left(1+\nu\right)\left[\dfrac{\dro^2}{\dro^2-1}\left(1-\dfrac{1}{\dr^2}\right) - \dfrac{\ln \dr}{\ln \dro}\right] - (1-b) \dPfi \dfrac{1}{\dro^2-1} \left(1-\dfrac{\dro^2}{\dr^2}\right)
\end{equation}

\begin{equation}\label{eq:anal23}
\bar{\sigma}_{\rm \theta}^{\rm m} = - \dfrac{1}{2} \dPfi \left(1+\nu\right)\left[\dfrac{\dro^2}{\dro^2-1}\left(1+\dfrac{1}{\dr^2}\right) - \dfrac{\ln \dr}{\ln \dro} + \dfrac{1}{\ln \dro} \dfrac{1-\nu}{1+\nu} \right] - (1-b) \dPfi \dfrac{1}{\dro^2-1} \left(1 - \dfrac{\dro^2}{\dr^2} \right)
\end{equation}

Inspection of equation (\ref{eq:anal21}) indicates that Biot's coefficient influences the displacement field strongly.

\end{document}